\documentclass[journal,twocolumn]{IEEEtran}
\usepackage{array}
\usepackage{amsmath}
\usepackage{lineno}
\usepackage{graphicx}  
\usepackage[caption=false, font=footnotesize]{subfig}
\usepackage{epstopdf}
\usepackage{bm}
\usepackage{amsthm,amsmath,amssymb}
\usepackage{algorithm}  
\usepackage{algorithmic}  
\usepackage{mathrsfs}
\usepackage{cite}
\usepackage{hyperref}
\usepackage{dsfont}
\usepackage{xcolor}
\usepackage {setspace}
\usepackage{stfloats}
\allowdisplaybreaks[4]
\newtheoremstyle{mystyle}{}{}{}{12pt}{}{:}{4pt}{\it{\thmname{#1}\thmnumber{ #2}}\thmnote{ (#3)}}
\theoremstyle{mystyle}
\newtheorem{theorem}{Theorem}
\newtheorem{lemma}{Lemma}

\renewcommand{\algorithmicrequire}{ \textbf{Input:}} 
\renewcommand{\algorithmicensure}{ \textbf{Output:}} 

\ifCLASSINFOpdf
\else
\fi
\begin{document}
	\bstctlcite{IEEEexample:BSTcontrol}  
	
	\title{A Tractable Approach to Massive Communication and Ubiquitous Connectivity in 6G Standardization}
	

	\author{Junyi~Jiang,~Wei~Chen,~\IEEEmembership{Senior~Member,~IEEE},~Xin~Guo,~\IEEEmembership{Member,~IEEE},~Shenghui Song,~\IEEEmembership{Senior~Member,~IEEE},\\Ying~Jun~(Angela)~Zhang,~\IEEEmembership{Fellow,~IEEE},~Zhu~Han,~\IEEEmembership{Fellow,~IEEE},~Merouane~Debbah,~\IEEEmembership{Fellow,~IEEE},~Khaled B. Letaief,~\IEEEmembership{Fellow,~IEEE}}

%


\maketitle
\begin{abstract}
	The full-scale 6G standardization has attracted considerable recent attention, especially since the first 3GPP-wide 6G workshop held in March 2025. To understand the practical and fundamental values of 6G and facilitate its standardization, it is crucial to explore the theoretical limits of spectrum, energy, and coverage efficiency considering practical hardware and signaling constraints. In this paper, we present a mean-field-approximation-based investigation on two out of six use case scenarios defined by IMT-2030, namely, massive communication and ubiquitous connectivity. Being aware of the limitation in interference cancellation owing to constrained cost and hardware complexity, we investigate the spectrum reuse architecture in both usage scenarios. We propose a tractable spectrum reuse with low signaling overhead consumed for channel estimation and channel state information (CSI) feedback. Our analysis indicates that the massive communication over cellular and device-to-device (D2D) networks can benefit from channel orthogonalization, while it is unnecessary to share the CSI of interfering links. Moreover, deploying relays or movable base stations, e.g. unmanned aerial vehicle,  yields substantial energy and spectrum gain for ubiquitous connectivity, despite introducing interference. As such, the mean-field-optimization-based evaluation is expected to positively impact 6G and NextG standardization in 3GPP and other standardization bodies.
\end{abstract}
\begin{IEEEkeywords}
	6G standardization, 3GPP, IMT-2030, massive communication, ubiquitous connectivity, spectrum sharing, power control, efficiency, coverage, practical constraints, CSI, mean-field approximation, performance limit, protocol design.
\end{IEEEkeywords}

\IEEEpeerreviewmaketitle
\section{Introduction}
Massive communications and ubiquitous connectivity are two of the most crucial usage scenarios of 6G networks to provide a seamless coverage for a broader number of entities from different applications. 3GPP has specified strict 6G standards that require $\mathrm{10^8}$ devices$\mathrm{/km^2}$ for connection density, $10$ Gbps$\mathrm{/m^2}$ for area traffic capacity, and 99\% coverage percentage to support innovative applications such as factory automation and smart city\cite{6G_you}. To meet the above requirements, we need to achieve scalable and efficient connectivity for a massive number of users and construct a hyper-connected network\cite{Hyper_connt}.

The support of massive communication has been widely investigated from different perspectives including both cellular networks and device-to-device (D2D) networks. For conventional massive cellular networks, the capacity of a massive access channel was first investigated in \cite{massive_capacity1} by considering a many-access channel (MnAC), where both the number of users and the code blocklength tend to infinity. While \cite{massive_capacity1} considered scenarios with known active transmitters, the case with randomly active transmitters was studied in \cite{massive_capacity2}, demonstrating that the cost of the activity identification is the same as the entropy of the activity probability \cite{massive_capacity3}. The extension to scenarios with multiple antenna was further investigated in \cite{massive_capacity_MIMO}. Massive access techniques have been widely discussed for massive cellular networks. In particular, the conventional orthogonal multiple access techniques struggle to support massive access because of the limited available spectrum \cite{NOMA_massive}. As a result, non-orthogonal multiple access (NOMA) has emerged as a promising solution as it can serve more than one user in the same resource block \cite{NOMA_poor}, \cite{NOMA_D}. In NOMA, interference is mitigated by superposition coding with weighted transmission power. However this approach leads to high computational complexity when a large number of devices\cite{PD_NOMA}. To address this issue, the devices are usually grouped into several clusters to reduce the computational complexity \cite{cluster1}, \cite{cluster2}. 

Massive D2D networks have attracted considerable attention for supporting the applications that require direct communication among devices \cite{MMTC_D2D}. In D2D networks, the key challenge is to find an appropriate scheduling strategy to meet the requirements of users such as delay and throughput. The scheduling mechanisms can be categorized to centralized scheduling and distributed scheduling, depending on whether there exists a scheduler knowing the global information of all D2D devices \cite{D2D_scheduling}. For centralized scheduling, an opportunistic subchannel scheduling algorithm was proposed in \cite{subchannel}  to maximize the average sum rate of the system. In this scheme the BS opportunistically chooses each user's transmission mode and allocates each user an exclusive subchannel according to channel conditions and individual user requirements. In \cite{D2D_EE}, a resource block is assigned to the link with highest ratio of transmitted bits to required energy, thereby maximizing the system.' energy efficiency. For distributed scheduling, a synchronous peer-to-peer wireless network architecture named FlashLinQ was proposed in \cite{FlashLinQ}, which enables efficient spatial resource allocation based on channel state information.      

To provide ubiquitous connectivity for a wide variety of applications without coverage holes, non-terrestrial networks are indispensable to enhance coverage and ensure connectivity for remote areas as a complement to traditional terrestrial networks (TN) \cite{NTN_survey}. Due to their high mobility, unmanned aerial vehicles (UAVs) play a vital role in non-terrestrial networks\cite{UAV_survey}. By mounting communication transceivers on UAVs, a swarm of UAVs can be deployed as flying base stations to construct an infrastructure-free network that provides reliable services to ground devices \cite{UAV_BS}. Compared to conventional base stations, UAV enabled BSs are more flexible, cost-effective and capable of establishing line-of-sight communication links \cite{UAV_prompt}. Therefore, the effective deployment of UAVs has been widely investigated with different objectives. In \cite{UAV_number}, \cite{UAV_number2}, a centralized deployment algorithm was proposed based on the position of ground devices to minimize the number UAVs required. \cite{UAV_cover} aimed to maximize the coverage of UAVs while minimizing transmit power. In \cite{UAV_capacity}, \cite{UAV_cover2}, UAVs were deployed as facilitators to enhance network capacity and coverage in the presence of fixed BSs. 

In order to improve the spectrum efficiency, tremendous spectrum sharing strategies have been proposed. However, due to the complex coupling mutual interference between the users, only simplified scenarios have been studied. For instance, \cite{MAPEL_high_SINR} considered the power allocation problem under a  
high signal-to-interference-plus-noise ratio (SINR) regime. \cite{capacity_gupta} investigated the scaling limit of ad hoc networks with the assumption that all nodes in the network transmit at the same rate. The problems become more complicated in case of large-scale networks owning to the exponentially increasing computational complexity with respect to the number of the users. Fortunately, the mean-field approximation, which was firstly applied to analyze the many-body physical system \cite{MF}, is a potential solution to simplify the power control for the system with massive users. In \cite{MF1}, \cite{MF2}, the mean-field approximation was applied in queuing systems to investigate the interaction between the objects. In \cite{MFG}, an energy-efficient velocity control algorithm for large-scale UAVs is obtained by formulating the optimal control problem as a mean field game (MFG). The MFG has also been applied in environment sensing problem for mobile crowd integrated sensing, communication, and computation systems to handle the complex interactions among large-scale devices and derive the optimal waveform precoding design \cite{MFG2}. In our previous works \cite{MF_JSAC}, \cite{MF_TWC}, the optimal cross-layer scheduling schemes were derived via mean-field approximation to minimize the average queuing delay and satisfy the hard delay constraints for massive devices.

In this paper, motivated by previously mentioned facts and challenges, we aim to analyze the fundamental limits of 6G networks under different usage scenarios. A tractable approach that is independent of the size of the network is proposed to analyze the throughput for 6G networks. Based on the proposed approach we discuss the effectiveness of spectrum sharing strategies such as channel orthogonalization and the requirements of user state information for 6G networks.
In particular, we consider both the networks for massive communication and ubiquitous connectivity. The objective is to investigate the throughput of networks with different network parameters including the channel uncertainty, the user intensity, and the power constraints. To tackle the complicated coupled interference of the users in the networks, the mean-field approximation is exploited to reduce the dimension of the throughput maximizing problems. A unified power control strategy is proposed for massive communication networks. We further analyze the scenario with the delay constraint where a MFG is constructed to obtain the optimal power allocation. For infrastructure-free network, an efficient routing strategy is firstly proposed, based on which we prove that the throughput of the network can be maximized by optimizing the transport capacity of corresponding single-hop network. Based on the obtained relationship between the network throughput and the network structure, we investigate the optimal design for 6G network. In particular, the effectiveness of channel orthogonalization like time division multiplexing (TDM) is analyzed and we compare the effect of user state information in terms of both its type and accuracy. We also provide the optimal network structure and power allocation for different requirements. The main contributions of this paper are as follows:   

\begin{itemize}
	\item We investigate the fundamental limits for different systems by analyzing their throughput with different device intensity and power constraints. These results can facilitate system design and standardization. 
	\item Based on the performance analysis of the 6G networks, the effectiveness of channel orthogonalization such as TDM is investigated and the accuracy requirements of the state information feedback from the users is provided for different usage scenarios.
	\item For massive network, We show that only the CSI between the source node and the destination of each link is needed for power allocation. This indicates that the distributed power control can be exploited in practical systems. For the usage scenario of ubiquitous connectivity, we show that the relay nodes can bring significant energy and spectrum gain for infrastructure-free networks based on the proposed routing strategy.  
	\item A tractable performance analysis approach that is independent of the size of the network is proposed for 6G networks based on the mean-field approximation. We prove that the mean-field approximation can effectively reduce the dimension of large-scale networks to obtain the optimal power allocation for the users in the network. 
\end{itemize}

The remainder of this paper is organized as follows. In Section \ref{sm}, we present the basic model for the 6G networks and formulate the primary throughput optimization problems. In Section \ref{SH}, we focus on the single-hop network and transform the original problem to a weighted throughput maximization problem with reduced dimension by mean-field approximation, based on which the optimal power control strategy is derived and then the performance limit of the network is investigated. In Section \ref{IF}, we first propose an efficient routing strategy for infrastructure-free network and the throughput of the network is investigated. Section \ref{NR} present the numerical results. Finally, we conclude this paper in section \ref{con}.  
\section{System Model}
\label{sm}
\begin{figure*}[ht]
	\centerline{\includegraphics[width=12.5cm]{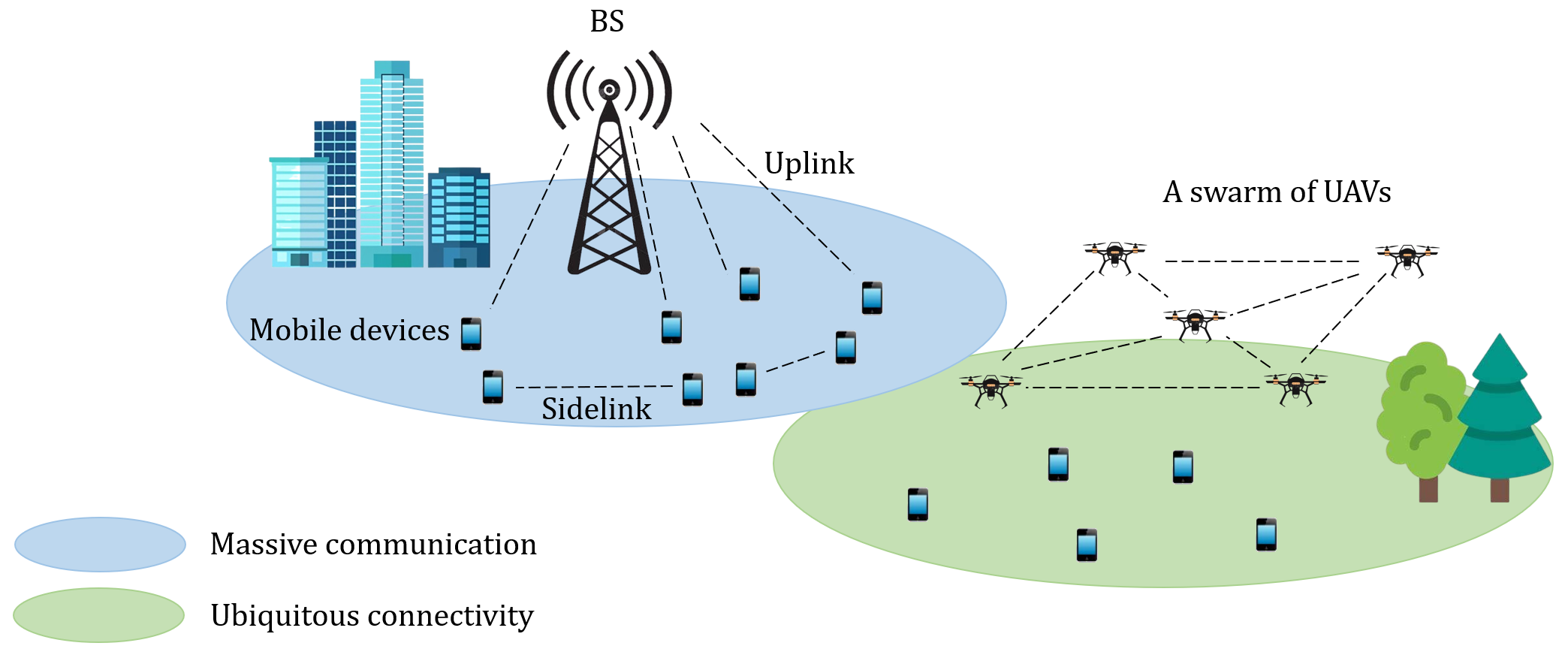}}
	\caption{Illustration of 6G networks for massive communication and ubiquitous connectivity}
	\label{sys_mod}
\end{figure*}
In this section, we present the network model for massive communication and ubiquitous connectivity. The original average transmission rate maximizing problems for the networks will also be formulated.
\subsection{Networks for Massive Communication}
As shown in Fig. \ref{sys_mod}, we consider a network consisting of a base station (BS) and $N_o$ users. We assume that the users are distributed according to homogeneous Poisson point process (PPP) of intensity $\lambda$. This indicates that for each small region with area $S$, the number of users in the region follows a Poisson distribution with parameter $\left\lfloor \lambda S\right\rfloor$ and for any two disjoint area $S_i$ and $S_j$, the number of users in two regions are independent\footnote{To investigate the performance limit of the networks, we assume that the devices are distributed according to homogeneous PPP in this paper. However, the proposed power control strategy is independent of the specific distribution of the devices and can be applied to practical systems according to the actual distribution.}. For each user, communication can occur via the uplink channel to the BS or directly through a sidelink channel to the target user. We denote by $T_i$ and $R_i$ the source node and the destination node (or BS) for the $i$th link, respectively. We denote by $G_{ij}$ the channel gain from $T_i$ to $R_j$, which is determined by both the path loss and small-scale fading. In this paper, we consider a discrete distance model. In particular, we denote by $d_{ij}\in\left\{d_0, 2d_0,...\right\}$ the distance between the $T_i$ and $R_j$, and we assume that the maximum transmission distance for each link is by $N_md_0$. Since all nodes are PPP distributed, the distribution of the distance between any two nodes is independently and identically distributed (i.i.d.). We define $\gamma_k=\mathrm{Pr}\left\{d_{ij}=kd_0\right\}$, which is equal to the probability that node $R_j$ locates in the ring with inner radius $(k-1)d_0$ and outer radius $kd_0$ of node $T_i$, i.e.,
\begin{equation}\label{gamma_a}
	\gamma_k=\frac{\pi d_0^2\left(k^2-(k-1)^2\right)}{\pi {N_m}^2d_0^2}=\frac{k^2-(k-1)^2}{{N_m}^2}.
\end{equation} 
Let $h_{ij}\in\left\{h_1, h_2,..., h_{N_h}\right\}$ denote the small scale fading factor between $T_i$ and $R_j$, where $N_h$ is the number of small-scale fading states. The small scale fading factor is assumed to be i.i.d. for all nodes. We denote by $\beta_k$ the probability that $\mathrm{Pr}\left\{h^c_{ij}=h_k\right\}$ for $1\le k\le N_h$ where $\sum_{k=1}^{N_h}\beta_k=1$. As a result, the channel gain between $T_i$ and $R_j$ is given by
\begin{equation}\label{PL}
	G_{ij}=h_{ij}f(d_{ij}),
\end{equation}
where $f(d_{ij})$ is the power attenuation function. In order to avoid the singularity at the origin, the bounded path-loss model (BPM) is applied. The power attenuation function of BPM is defined as $f(d)=(1+d)^{-\alpha}$. 

Let $p_i$ denote the transmit power for $T_i$. The signal received by $R_i$ is given by
\begin{equation}
	y_i=\sum_{j=1}^{N_o}\sqrt{G_{ji}p_j}+z,
\end{equation} 
where $x_i$ is the signal transmitted by $T_i$ and $z$ is Additive White Gaussian Noise (AWGN) with power $n$. Thus, the maximum transmission rate of link $i$ is
\begin{equation}
	r_i=\log_{2}\left(1+\frac{G_{ii}p_i}{\sum_{j\ne i}G_{ji}p_j+n}\right),
\end{equation}
We aim to maximize the average transmit rate while ensuring that both the maximum power and minimum rate constraints are satisfied. The problem can be formulated as
\begin{subequations}\label{mc_ori}
	\begin{align}
		\max_{\boldsymbol{p}} \frac{1}{N_o}\sum_{i=1}^{N_o}\log_{2}&\left(1+\frac{G_{ij}p_i}{\sum_{j\ne i}G_{ji}p_j+n}\right)\\[4mm]
		s.t.\quad\log_{2}\Bigg(1+&\frac{G_{ii}p_i}{\sum_{j\ne i}G_{ji}p_j+n}\Bigg)\ge R_{\text{min}},\\[4mm]
		&p_i\le p_{\text{max}}.
	\end{align}
\end{subequations} 
where $\boldsymbol{p}=(p_i, 1\le i\le N_o)$ is the power vector, and $r_\mathrm{min}$ and $p_\mathrm{max}$ are the minimum rate and maximum power, respectively.

\subsection{Networks for Ubiquitous Connectivity}
For ubiquitous connectivity networks, we consider the systems that operate independently of the fixed infrastructure. In these systems, users are allowed to operate as relay nodes, forwarding data between one another. We refer to this system as an infrastructure-free network. Similar to the previous networks, we also consider PPP distributed nodes in the area $S$ with intensity $\lambda$. The number of the nodes is given by $N_o=\left\lfloor \lambda S\right\rfloor$. For each user, the destination node is randomly chosen within the range $N_md_0$. Hence the transmit distance for each link satisfies $d_{ii}\in\left\{d_0, 2d_0,..., N_md_0\right\}$. The destination node of the $i$th link with the distribution of $d_{ii}$ is given by (\ref{gamma_a}). The channel gain is determined by both small-scale fading and path loss which is given by (\ref{PL}).

For the infrastructure-free network, the average rate for each link is defined as the the ratio of the transmitted bits to the required time, and the average rate of the network is defined as the average rate of all links. Let $r_i$ denote the average rate of the $i$th link. Our objective is to develop an effective routing strategy and the optimal power allocation to maximize the average rate of the network while adhering to the imposed power constraints.
\begin{subequations}
	\begin{align}
		\max_{\boldsymbol{p}}\ &\sum_{i=1}^{N_o}r_i(\boldsymbol{p})\\
		s.t. \quad&r_i(\boldsymbol{p})\ge R_{\mathrm{min}}\\
		\quad&p_i\le p_{\mathrm{max}}.
	\end{align}
\end{subequations}
In the following sections, we employ a mean-field approximation to reduce the dimensionality of the problem and obtain the optimal power control strategy for the networks.

\section{The capacity for Massive Communication Networks}
\label{SH}
In this section, we aim to simplify Problem (\ref{mc_ori}) by reducing the dimensionality of the optimization problem via mean-field approximation. The simplified optimization problem is then utilized to determine the optimal power allocation and the corresponding network capacity.

\subsection{Mean-Field Approximation }

For massive communication networks, the data rate for each link is determined by both the link's channel gain and the interference power at its destination node. As a result, transmit power is allocated based on the signal and interference power. Therefore, we first derive the distribution of the link channel gain and interference power. 

The distribution of channel gain $G_{ii}$ is determined by the joint distribution of $d_{ii}$ and $h_{ii}$. We define $g_{kl}=h_kf(ld_0)$, for $k\in\left\{1,2,...,N_h\right\}$ and  $l\in\left\{1,2,...,N_m\right\}$. Let $\theta_{kl}$ denote the probability that the channel gain equals to $g_{kl}$ i.e. $\theta_{kl}=\mathrm{Pr}\left\{G_{ii}=g_{kl}\right\}$ for $ i\in\left\{1,2,...,N_o\right\}, k\in\left\{1,2,...,N_h\right\}$ and $l\in\left\{1,2,...,N_m\right\}$, which is given by
\begin{equation}
	\theta_{kl}=\beta_k\gamma_l=\beta_k\frac{l^2-(l-1)^2}{{N^c_m}^2}.
\end{equation}
To simplify the notation, we sort $g_{kl}$ in a descending order. The sorted element is denoted by $g_m$, $m\in\left\{1,2, ...,N_g\right\}$, where $N_g=N_hN_m$ denotes the number of channel gain state. Hence, we have $g_1\ge g_2\ge...\ge g_{N_g}$. We denote the mapping from the two dimensional subscript $kl$ to the new subscript $m$ by $\phi$, i.e. $m=\phi\left(k,l\right)$. We also update the subscript of $\theta_{kl}$ according to $\phi$ and denote the updated result by $\theta_m$, i.e. $\theta_m=\mathrm{Pr}\left\{G_{ii}=g_m\right\}$. 

For the interference power, we consider a circular region of radius $r_o$ centered on the destination node $R_i$. The total number of nodes in the region is given by $\left\lceil\lambda\pi r_o^2\right\rceil$. Let $I_{r_o}$ denote the interference power at $R_i$. convergence of $I_{r_o}$ is presented in the following lemma.  
\begin{lemma}
	\label{lemma1}
	When $\alpha>2$, the average interference power at the destination node converges and can be approximated by $I_{r_0}$ when $r_0$ is sufficiently large. 
\end{lemma} 
\begin{IEEEproof}
	For the destination node $R_j$, we denote by $M=\left\lceil \lambda\pi r_o^2\right\rceil$ the number of interference nodes within a region of radius $r_o$. The average interference power is given by 
	\begin{equation}\label{lemma1.1}
		\mathbb{E}[I_{r_o}]=\mathbb{E}\left[\sum_{i=1}^{M}I_{r_o}^{\left(i\right)}\right],
	\end{equation}
	where $I_{r_o}^{\left(i\right)}$ is the interference power from the $i$th node within the considered region. Since the interference from each node is independent, we have
	\begin{equation}\label{lemma1.2}
		\mathbb{E}\left[\sum_{i=1}^{M}I_{r_o}^{\left(i\right)}\right]=\sum_{i=1}^{M}\mathbb{E}\left[I_{r_o}^{\left(i\right)}\right].
	\end{equation}
	For interference from the $i$th node, we have
	\begin{equation}
		\mathbb{E}\left[I_{r_o}^{\left(i\right)}\right]=\mathbb{E}\left[p_i h_{ij}f\left(d_{ij}\right)\right],
	\end{equation}
	where $p_i$ denotes the transmit power of the $i$th node, and $h_{ij}$ and $f(d_{ij})$ represent small-scale fading and path-loss between the $i$th node and destination node $R_j$, respectively. Because the power of the $i$th node is allocated based on the channel gain and the total interference power of it, $p_i$, $h_{ij}$ and $f\left(d_{ij}\right)$ can be considered to be independent. We have
	\begin{equation}
		\label{lemma1.11}
		\begin{split}
			\mathbb{E}\left[p_i h_{ij}f\left(d_{ij}\right)\right]&=\mathbb{E}\left[p_i\right]\mathbb{E}\left[h_{ij}\right]\mathbb{E}\left[f\left(d_{ij}\right)\right]\\ 
			&=\bar{p}\bar{h}\int_{0}^{r_o}f(r)f_D(r)\mathrm{d}r.
		\end{split}
	\end{equation}
	where $\bar{p}$ and $\bar{h}$ are the average power and average small-scale fading factor, respectively. $f_D(r)$ is the probability density function of $d_{ij}$, which is given by $f_D(r)=\frac{2}{r_o^2}r$. Substituting it and $f(r)=\left(1+r\right)^{-\alpha}$ into (\ref{lemma1.11}) we have
	\begin{equation}\label{lemma1.3}
		\begin{split}
			\mathbb{E}\left[p_i h_{ij}f\left(d_{ij}\right)\right]&=\bar{p}\bar{h}\int_{0}^{r_o}\left(1+r\right)^{-\alpha}\frac{2}{r_o^2}r\mathrm{d}r\\
			&=\frac{2}{r_o^2}\bar{p}\bar{h}\left(\frac{\left(1+r_o\right)^{1-\alpha}\left(1+\left(\alpha-1\right)r_o\right)-1}{\left(2-\alpha\right)\left(\alpha-1\right)}\right).
		\end{split}
	\end{equation}
	Substituting  \eqref{lemma1.2} and \eqref{lemma1.3} into  \eqref{lemma1.1}, we have
	\begin{equation}
		\begin{split}
			\mathbb{E}[I_{r_o}]&=\sum_{1}^{M}\frac{2}{r_o^2}\bar{p}\bar{h}\left(\frac{\left(1+r_o\right)^{1-\alpha}\left(1+\left(\alpha-1\right)r_o\right)-1}{\left(2-\alpha\right)\left(\alpha-1\right)}\right)\\
			&=\left\lceil \lambda\pi r_o^2\right\rceil\frac{2}{r_o^2}\bar{p}\bar{h}\left(\frac{\left(1+r_o\right)^{1-\alpha}\left(1+\left(\alpha-1\right)r_o\right)-1}{\left(2-\alpha\right)\left(\alpha-1\right)}\right).
		\end{split}
	\end{equation}
	When $\alpha>2$, we have
	\begin{equation}
		\mathbb{E}[I_{r_o}]<\left(\lambda\pi+\frac{1}{r_o^2}\right)\left(\frac{\bar{p}\bar{h}}{(\alpha-2)(\alpha-1)}\right),
	\end{equation}
	which shows that the average interference power is converge, the proof is completed. 
\end{IEEEproof}
According to Lemma \ref{lemma1}, it is reasonable to approximate the total interference power as $I_{r_o}$ when $r_o$ is sufficiently large. In this paper, we denote by $d^I_{m}=N^I_md_0$  the maximum distance between the source node $T_i$ and destination node $R_j$ when $i\ne j$, i.e. $r_o=d^I_m$. Consequently the total number of interference node is given by $N_i=\left\lceil\lambda\pi\left(N^I_md_0\right)^2\right\rceil$.     

Let $\gamma^I_k$ denote the probability that the distance between $T_i$ and $R_j$ is $kd_0$, i.e. $\gamma^I_k=\mathrm{Pr}\left\{d_{ij}=kd_0\right\}, 1\le k\le {N^I_m}, i\ne j$. Similar to Eq. (\ref{gamma_a}), this probability is given by 
\begin{equation}\label{gamma_a_I}
	\gamma^I_a=\frac{a^2-(a-1)^2}{{N^I_m}^2}.
\end{equation}
We define the interference channel gain as $g^I_{kl}=h_kf(ld_0)$ and denote by $\theta^I_{kl}=\mathrm{Pr}\left\{G_{ij}=g^I_{kl}\right\}$ the distribution of interference gain, which is given by 
\begin{equation}
	\theta_{kl}^I=\beta_k\frac{a^2-(a-1)^2}{{N^I_m}^2}.
\end{equation}
Similar simplification is applied to $g_{kl}^I$ and $\theta_{kl}^I$. As a result, we obtain $g_m^I$ and $\theta_{m}^I$ for $m\in\left\{1,2,...,N_g^I\right\}$, representing the possible values of the interference channel gain and their corresponding probabilities, where $N_g^I=N_hN^I_m$ is the number of possible values of $g_m^I$.

We then define a vector $\boldsymbol{a}^j\in\mathbb{R}^{1\times N^I_g}$ to describe the channel gain between the interference nodes and destination node $R_j$. The $k$th element of $\boldsymbol{a}^j$ is the number of nodes for which the channel gain between them and $R_j$ is equal to $g^I_k$. That is $a^j_k=n$ means that there are $n$ nodes satisfying $G_{ij}=g_k^I, i\ne j$ for $R_j$.  Meanwhile, we denote by $p_{a_i}^n$ the probability that $a_i^j=n$, which is given by 
\begin{equation}
	p_{a_i}^n={\theta^I_i}^n\left(1-\theta^I_i\right)^{N_i-n}C_{N_i}^n,
\end{equation}
where $C_{N_i}^n$ is the binomial coefficient.  The complete interference matrix that including the distribution of interference channel gain for each destination node is denoted by $\boldsymbol{A}=[\boldsymbol{a}^1\;\boldsymbol{a}^2...\;\boldsymbol{a}^{N_o^c}]$. Thus, the network's interference channel gain information is encapsulated $\boldsymbol{A}$ and can be exploited for power allocation. However, because of the its high dimensionality, this problem is too complex to solve directly. Therefore, we aim to reduce the dimension of the problem by categorizing the nodes into different group according to their interference channel gains as follows.  

Recall that the interference channel gains $g^I_m$ are sorted in a descending order, i.e., $g^I_1\ge g^I_1\ge...\ge g^I_{N^I_g}$. This ordering indicates that for destination $R_j$, the power of the interference signal from the node with channel gain $g^I_m$ is small when $m$ is large. As a result, we can primarily focus on the first $N_a$ elements of $\boldsymbol{a}$. To further reduce the problem's dimensionality, we partition all possible values of $a_k$ into $N_c$ intervals that have equal probabilities. We then and quantize these values using the centroid of each interval.  Let $\mathcal{I}_{kl}$ denote the set of possible value in the $l$th interval for the $k$th element of $\boldsymbol{a}$. This set is determined such 
\begin{equation}
	\sum_{i\in\mathcal{I}_{kl}}p_{a_k}^i=\frac{1}{N_c},
\end{equation}
for $\forall k\in\left\{1, 2,..., N_g^I\right\}$ and $\forall l\in\left\{1, 2,..., N_c\right\}$. The centroid of $\mathcal{I}_{kl}$ is denoted by $\bar{a}_{ij}$, which is given by
\begin{equation}
	\bar{a}_{kl}=N_c\sum_{i\in\mathcal{I}_{kl}}ip_{a_k}^i.
\end{equation}
We then categorize the interference channel gain vector $\boldsymbol{a}$ into $N_I$ groups according to its first $N_a$ elements, where $N_I=N_c^{N_a}$. Let $\mathcal{A}_k$ denote the $k$th group. For each group $\mathcal{A}_k$ we define vector $\boldsymbol{u}^k\in\mathbb{R}^{1\times N_a}$ to indicate the intervals to which the elements of the interference channel gain vectors belong . In particular, for $\boldsymbol{a}\in\mathcal{A}_k$, the interval corresponding to the $i$th element of $\boldsymbol{a}$ is given by $u^k_i$. Besides, the mapping between $\boldsymbol{u}^k$ and $k$ is
\begin{equation}
	\sum_{i=1}^{N_a}u^k_i(N_c)^{i-1}=k.
\end{equation}
Recall that the total number of interfering nodes is given by $N_i$. For users in the $k$th group, the number of interfering node with channel gain in $\left\{g^I_i|1\le i\le N_a\right\}$ is $\sum_{i=1}^{N_a}\bar{a}_{iu^k_i}$. We further denote by $N_r^k$ the number of remaining nodes with channel of $\left\{g_i^I|N_a<i\le N_g^I\right\}$. It is given by $N_r^k=N_i-\sum_{i=1}^{N_a}\bar{a}_{iu^k_i}$. Since $N_g^I$ is sufficiently large, we assume that the first $N_a$ elements in $\boldsymbol{a}$ are independent. Therefore, the probability that $\boldsymbol{a}$ belongs to each group are the same for all groups, and is given by $\xi=\frac{1}{N_I}$. 

In order to present the mean-field approximation of Problem (\ref{mc_ori}), we first define the weight matrix $\boldsymbol{W}$, posterior matrices $\boldsymbol{Q}^1$ and $\boldsymbol{Q}^2$ and the posterior vector $\boldsymbol{q}$ as follows. The element in the $i$th row and the $j$th column of $\boldsymbol{W}$ is defined as the probability that the destination node $R_k$ is in the $i$th interference group and the channel gain between it and the destination node $T_l$ is given by $g_j$. This probability is given by $W_{ij}=\xi\theta_j$. The element in the $i$th row and the $j$th column of $\boldsymbol{Q}^1$ is defined as the conditional probability that given the interference gain between the node and the destination node is $g^I_i$, the node is in set $\mathcal{A}_j$. It is given by 
\begin{equation}
	Q^1_{ij}=\frac{\xi Q^2_{ji}}{\sum_{k=1}^{N_I}\xi Q^2_{jk}}=\frac{ Q^2_{ji}}{\sum_{k=1}^{N_I} Q^2_{jk}},
\end{equation}  
where the element in the $i$th row and the $j$th column of $\boldsymbol{Q}^2$ is defined as the probability that, given the node is in $\mathcal{A}_i$, the interference channel gain between the node and the destination node is $g^I_j$. This probability is given by
\begin{equation}
	Q^2_{ij}=\frac{\bar{a}_{ju^i_j}}{\sum_{k=1}^{N_a}\bar{a}_{ku^i_k}}.
\end{equation} 
The $i$th element of $\boldsymbol{q}$ is defined the conditional probability that the interference node is in $\mathcal{A}_k$ given the channel gain between the node and the destination is in $\left\{g^I_i|1+N_a\le i\le N_g^I\right\}$, which is given by
\begin{equation}
	q_i=\frac{{N}_r^i}{\sum_{j=N_a+1}^{N_g}{N}_r^j}.
\end{equation} 
Below is a formal statement of the mean-field approximation of Problem (\ref{mc_ori}) using the variables defined above:
\begin{theorem}
	\label{theorem 1}
	The mean-field approximation of Problem (\ref{mc_ori}) is given by the following WTM problem, i.e.,
	\begin{subequations}\label{TH1}
		\begin{align}
			\max_{\mathcal{P}}\quad&\sum_{i=1}^{N_I}\sum_{j=1}^{N_g}W_{ij}\log_{2}\left(1+\frac{\mathcal{P}_{ij}g_j}{I_i+n}\right)\\\nonumber
			s.t.\quad \ &I_i =\sum_{k=1}^{N_a}\sum_{l=1}^{N_I}\sum_{m=1}^{N_g}\bar{a}_{k\boldsymbol{u}_i(k)}Q^1_{kl}\theta_m{\mathcal{P}_{lm}}g^I_k\\
			& \ \ +\sum_{k=N_a+1}^{N_g^I}\sum_{l=1}^{N_I}\sum_{m=1}^{N_g}N_r^i\theta^I_kq_l\mathcal{P}_{lm}g_k^I, \\
			&\log_{2}\left(1+\frac{\mathcal{P}_{ij}g_j}{I_i+n}\right)\ge R_{\mathrm{min}},\\
			&\mathcal{P}_{ij}\le p_{\mathrm{max}},
		\end{align}
	\end{subequations}
	where $\mathcal{P}_{ij}$ is the transmit power of links of which the destination node is in $\mathcal{A}_i$ and the channel gain is equal to $g_j$.
\end{theorem}
\begin{IEEEproof}
	We first prove that the mean-field approximation for the objective function in  (\ref{mc_ori}a) is given by  (\ref{TH1}a). According to the law of large numbers, approximately $N^c_o\xi$ links have their destination nodes are in each $i$th interference channel gain group $\mathcal{A}_i$. Since the interference power for each link is independent of its channel gain, we denote by $I_i$ the interference power for the links with destination nodes in $\mathcal{A}_i$. Besides, in each group, the number of links with channel gain $g_j$ is given by $N^c_o\xi\theta_j$. Hence the information rate for these links is expressed as
	\begin{equation}
		R_{ij} = \log_{2}\left(1+\frac{\mathcal{P}_{ij}g_j}{I_i+n}\right).
	\end{equation}  
	The average rate of the network is then
	\begin{equation}\label{proof_target}
		R_{\mathrm{ave}}=\frac{1}{N_o}\sum_{i=1}^{N_I}\sum_{j=1}^{N_g}N_o\xi\theta_j\log_{2}\left(1+\frac{\mathcal{P}_{ij}g_j}{I_i+n}\right).
	\end{equation}
	Replacing $\beta\theta_j$ by $\omega_{ij}$, we have obtained (\ref{TH1}a).
	
	For the interference power of a destination node in the $i$th group, we first calculate the interference power contributed by nodes with a specific $g_k^I$, and then sum up these contributions to obtain the total interference power. Specifically, for a destination node $R_i\in \mathcal{A}_i$, let $I_i ^{(k)}$ denote the total interference power from nodes with interference gain $g^I_k$ for $1\le k\le N_a$. Given the interference channel gain between the interference node and the destination node, the probability that the interference node is in group $\mathcal{A}_l$ is given by $Q^1_{kl}$. Moreover, the probability that the channel gain of the interference node is $g_m$ is $\theta_m$. According to the law of large numbers, the number of nodes with interference channel gain $g_k^I$ corresponding to the destination node $R_i$ and the transmission power $\mathcal{P}_{lm}$ can be approximated as $\bar{a}_{k\boldsymbol{u}_i(k)}P_{kl}^1\theta_m$. As a result, the interference power $I_i^{(k)}$ is given by 
	\begin{equation}
		I_i^{(k)}=\sum_{l=1}^{N_I}\sum_{m=1}^{N_g}\bar{a}_{k\boldsymbol{u}_i(k)}Q^1_{kl}\theta_m{\mathcal{P}_{lm}}g^I_k,
	\end{equation}
	for $1\le k\le N_a$. Similarly, for the interference nodes with interference channel gains in $\left\{g_i^I|N_a<i\le {N^I_m}\right\}$, the law of large numbers implies that, the number of the nodes with interference gain $g_k^I$ and transmission power $\mathcal{P}_{lm}$ is approximated by $N_r^i\theta^I_kp_l$. As a result, we have
	\begin{equation}
		I_i^{(k)}=\sum_{l=1}^{N_I}\sum_{m=1}^{N_g}N_r^i\theta^I_kq_l\mathcal{P}_{lm}g_k^I,
	\end{equation}
	for $N_a<k\le N_g^I$. Summing up $I_i^{(k)}$, we have 
	\begin{equation}\label{proof_interference}
		\begin{split}
			I_i&=\sum_{k=1}^{N_g^I}I_i^{(k)}\\
			&=\sum_{k=1}^{N_a}\sum_{l=1}^{N_I}\sum_{m=1}^{N_g}\bar{a}_{k\boldsymbol{u}_i(k)}Q^1_{kl}\theta_m{\mathcal{P}_{lm}}g^I_k\\
			&+\sum_{k=N_a+1}^{N_g^I}\sum_{l=1}^{N_I}\sum_{m=1}^{N_g}N_r^i\theta^I_kq_l\mathcal{P}_{lm}g_k^I.
		\end{split}
	\end{equation}
	Combining  (\ref{proof_target}) and (\ref{proof_interference}). the proof is completed. 
\end{IEEEproof}
Theorem \ref{theorem 1} indicates that by mean-field approximation, the dimension of Problem (\ref{mc_ori}) can be reduced from $N^c_o$ to $N_I\times N_g$. In large-scale networks, the total number of links $N^c_o$ grows rapidly with network density. However, the number of link channel states and interference channel states are independent of the network size. As a result, by mean-field approximation, the power control problem in large-scale wireless network can be simplified as a finite dimension problem.

%

%

\subsection{MAPEL algorithm}
In this subsection, we derive the efficient algorithm to solve Problem (\ref{TH1}). To simplify the problem and reduce its dimensionality, we first consolidate the groups of links. Specifically, we categorize the links whose destination nodes belong to $\mathcal{A}_i$ and whose channel gain is $g_j$ into a new group $\tilde{\mathcal{A}}_k$ according to the following mapping:
\begin{equation}
	k=\phi(i,j)=j+(i-1)N_g,
\end{equation}
for $ i\in\left\{1,...,N_I\right\}, j\in\left\{1,...,N_g\right\}$ and $ k\in\left\{1,...,N_t\right\}$, where $N_t=N_IN_g$. Besides, we denote by $\phi^{-1}_1(k)$ and $\phi^{-1}_2(k)$ the inverse mapping from $k$ to $i$ and $j$, respectively. Let  $\tilde{g}_k=g_{\phi^{-1}_2(k)}$, $\tilde{\theta}_k=\theta_{\phi^{-1}_1(k)}$ $\tilde{\boldsymbol{u}}_k=\boldsymbol{u}_{\phi^{-1}_1(k)}$ and $\tilde{N}_r^k=\tilde{N}_r^{\phi^{-1}_{1}(k)}$ for $\forall i\in\left\{1,...,N_t\right\}$. we define the new posterior probability matrix $\tilde{\boldsymbol{Q}}^1$ whose $(i,j)$th element is the probability that given the channel gain between the destination node and the interference node is $g^I_i$, the interference node is in $\tilde{\mathcal{A}}_j$. This probability is given by
\begin{equation}
	\tilde{Q}^1_{ij}=\frac{ \tilde{Q}^2_{ji}\tilde{\theta}_{j}}{\sum_{k=1}^{N_t}\tilde{Q}^2_{ki}\tilde{\theta}_{k}},
\end{equation}
where $\tilde{\boldsymbol{Q}}^2$ is the probability matrix whose $(i,j)$th element is the probability, that given the node in $\tilde{\mathcal{A}}_i$ the interference gain between it and the destination node is $g^I_j$, which is given by
\begin{equation}
	\tilde{Q}^2_{ij}=\frac{\bar{a}_{j\tilde{\boldsymbol{u}}_{i}(j)}}{\sum_{k=1}^{N_a}\bar{a}_{k\tilde{\boldsymbol{u}}_{i}(k)}}.
\end{equation}
Let $\tilde{\boldsymbol{q}}$ denote the probability vector whose $i$th element is the probability that the node is in $\tilde{\mathcal{A}}_i$ when the channel gain between the node and the destination is $\left\{g^I_{k}|N_a<k\le N^I_g\right\}$. This probability is given by
\begin{equation}
	\tilde{q}_i=\frac{\tilde{N}_r^i}{\sum_{j=1}^{N_t}\tilde{N}_r^j}.
\end{equation} 
Given the variables defined above, Problem (\ref{TH1}) is simplified as 
\begin{subequations}\label{MF_SH2}
	\begin{align}
		\max_{\tilde{\boldsymbol{p}}}\quad&\sum_{i=1}^{N_t}\tilde{\omega}_{i}\log_{2}\left(1+\frac{\tilde{p}_i\tilde{g}_i}{\tilde{I}_i+n}\right)\\\nonumber
		s.t.\quad \ &\tilde{I}_i =\sum_{j=1}^{N_a}\sum_{k=1}^{N_t}\bar{a}_{j\tilde{\boldsymbol{u}}_i(j)}\tilde{Q}^1_{jk}\tilde{p}_kg^I_j\\
		& \ \ +\sum_{j=N_a+1}^{N_g^I}\sum_{k=1}^{N_t}\tilde{N}_r^i\theta^I_k\tilde{q}_k\tilde{p}_kg_j^I, \\
		&\log_{2}\left(1+\frac{\tilde{p}_i\tilde{g}_i}{\tilde{I}_i+n}\right)\ge {R}_{\mathrm{min}},\\
		&\tilde{p}_i\le{p}_{\mathrm{max}}.
	\end{align}
\end{subequations}

According to Problem (\ref{MF_SH2}), the following equivalent interference gain matrix $\tilde{\boldsymbol{G}}$ whose $(i,j)$th entry is the equivalent interference gain from link in group $i$ to the link in group $j$. Specifically, $\tilde{G}_{i,j}$ is given by 
\begin{equation}\label{I1}
	\tilde{G}_{ij}=\sum_{k=1}^{Na}\bar{a}_{k\tilde{u}^i_j}\tilde{Q}^1_{kj}g^I_k+\sum_{k=N_a+1}^{N_g^I}\tilde{N}_r^i\theta_k^I\tilde{q}_jg_k^I.
\end{equation}
As a result, the interference experienced by the links in the $i$th group is given by
\begin{equation}\label{I2}
	I_i=\sum_{j=1}^{N_t}\tilde{p}_j\tilde{G}_{ji}.
\end{equation}
Because of the mutual interference between the links, there may not exist a feasible solution to Problem (\ref{MF_SH2}). We first present the algorithm to check the feasibility of the problem. Let $\gamma_{\mathrm{min}}$ denote the minimum SINR to achieve $R_{\mathrm{min}}$. This is calculated as $\gamma_{\mathrm{min}}=2^{{R}_{\mathrm{min}}-1}$. We then construct the feasibility check matrix $\boldsymbol{F}$ whose $(i,j)$th element is defined as
\begin{equation}\label{F_matrix}
	F_{ij}=\frac{\gamma_{\mathrm{min}}\tilde{G}_{ji}}{\tilde{g_i}}.
\end{equation} 
According to \cite{MC_network}, the problem is infeasible if the maximum eigenvalue of $F$ is greater than 1. Otherwise, a feasible power allocation is given by 
\begin{equation}\label{feasibility}
	\check{\boldsymbol{p}}=\left(I-F\right)^{-1}\boldsymbol{b},
\end{equation}  
where $\boldsymbol{I}$ is the $N_t\times N_t$ identity matrix and $\boldsymbol{b}$ is a $M\times 1$ vector which is given by
\begin{equation}
	b_i=\frac{\gamma_{\mathrm{min}}n}{\tilde{g}_i}.
\end{equation}
If $\check{\boldsymbol{p}}$ satisfies $0\le\check{p}_i\le{p}_{\mathrm{max}}$, Problem (\ref{MF_SH2}) is feasible.

In the following, we solve Problem (\ref{MF_SH2}) by MAPEL algorithm which is the first algorithm to achieve the global optimal solution of the WTM problems in the general SINR scheme. Before this, we first show that Problem (\ref{MF_SH2}) belong to a special class of Generalized Fractional Linear Programming (GFLP) in the following lemma \cite{MAPEL}. 
\begin{lemma}
	\label{lemma2}
	Problem (\ref{MF_SH2}) belongs to the class of Multiplicative Fractional Linear Programming (MFLP). 
\end{lemma}
\begin{IEEEproof}
	Let $\tilde{f}_i(\tilde{\boldsymbol{p}})=\tilde{\boldsymbol{p}}_i\tilde{g}_i+I_i+n$ and $\hat{f}_i(\tilde{\boldsymbol{p}})=I_i+n$. Then the objective of Problem (\ref{MF_SH2}) can be written as
	\begin{equation}
		\max_{\tilde{\boldsymbol{p}}}\quad\prod_{i=1}^{N_t}\left(\frac{\tilde{f}_i(\tilde{\boldsymbol{p}})}{\hat{f}_i(\tilde{\boldsymbol{p}})}\right)^{\tilde{\omega}_i},
	\end{equation}
	which is the product of exponentiated linear fractional functions and is increasing on $\mathcal{R}_{+}^{N_t}$. Meanwhile, the feasible set of Problem (\ref{MF_SH2}) is given by 
	\begin{equation}
		\mathcal{F}=\left\{\tilde{\boldsymbol{p}}|0\le \tilde{\boldsymbol{p}}_i\le p_{\mathrm{max}}, \frac{\tilde{f}_i(\tilde{\boldsymbol{p}})}{\hat{f}_i(\tilde{\boldsymbol{p}})}\ge 2^{
			{R}_{\mathrm{min}}}\right\}.
	\end{equation} 
	Based on the definition of GFLP \cite{MAPEL}, Problem (\ref{MF_SH2}) is a GFLP. This completes the proof.
\end{IEEEproof}
In the following, we present the MAPEL algorithm to solve Problem (\ref{MF_SH2}). Let us  define vector $\boldsymbol{z}$, whose $i$th element is  $z_i=\frac{\tilde{f}_i(\tilde{\boldsymbol{p}})}{\hat{f}_i(\tilde{\boldsymbol{p}})}$. We further denote the feasible region of $\boldsymbol{z}$ by $\mathcal{Z}$ as
\begin{equation}
	\mathcal{Z}=\left\{\boldsymbol{z}|0\le\boldsymbol{z}_i\le\frac{\tilde{f}_i(\tilde{\boldsymbol{p}})}{\hat{f}_i(\tilde{\boldsymbol{p}})},\forall i\in\left\{1,...,N_t\right\}, \tilde{\boldsymbol{p}}\in\mathcal{F}\right\}.
\end{equation} 
\begin{algorithm}[t]
	\caption{MAPEL Algorithm}
	\label{MAPEL}
	\renewcommand{\algorithmicrequire}{\textbf{Input:}}
	\renewcommand{\algorithmicensure}{\textbf{Output:}}
	\begin{algorithmic}[1]
		\STATE\textbf{Initialization}: Check the feasibility of the constraints by  (\ref{feasibility}) and (\ref{MF_SH2}d). If the constraints are infeasible, terminate the algorithm. Otherwise, choose the approximation factor $\delta_0>0$, and let $j=1$.
		\REPEAT
		\IF{$j=1$}
		\STATE Construct the initial vertex set $\mathcal{V}_1=\left\{\boldsymbol{v}^{(1)}\right\}$, the $i$th element of which is given by 
		\begin{equation}\nonumber
			v^{(1)}_i=\max_{\tilde{\boldsymbol{p}}\in\left[\boldsymbol{0},\boldsymbol{\tilde{p}}_{\mathrm{max}}\right]}\frac{\tilde{f}_i(\tilde{\boldsymbol{p}})}{\hat{f}_i(\tilde{\boldsymbol{p}})}=1+\frac{\tilde{g}_ip_{\mathrm{max}}}{n},\forall i\in\left\{1,...,N_t\right\}
		\end{equation}
		\ELSE
		\STATE Construct a new vertex set $\mathcal{V}_j$ by replacing $\boldsymbol{z}_{j-1}$ in $\mathcal{V}_{j-1}$ with $N_t$ new vertices $\left(\boldsymbol{v}^{(j)}_1,...,\boldsymbol{v}^{(j)}_{N_t}\right)$, where $\boldsymbol{v}^{(j)}_m=\boldsymbol{z}^{(j-1)}-\left(\boldsymbol{z}^{(j-1)}_m-\boldsymbol{\pi}^{(j-1)}_{m}\right)\boldsymbol{e}_m$.
		\STATE Remove the vertex $\boldsymbol{v}$ from $\mathcal{V}_j$ if $\boldsymbol{v}$ satisfies $\left\{\boldsymbol{v}|\boldsymbol{v}\prec\tilde{\boldsymbol{v}},\tilde{\boldsymbol{v}}\in\mathcal{V}_j\right\}$  
		\ENDIF
		\STATE Find $\boldsymbol{z_j}$ that maximize the objective function   (\ref{MF_SH2}a) over set $\mathcal{V}_j\cap\Theta$
		\begin{equation}\nonumber
			\boldsymbol{z}^{(j)}=\arg\max\left\{\prod_{i=1}^{N_t}\left(v_i\right)^{\tilde{\omega}_i}|\boldsymbol{v}\in\mathcal{V}_j\cap\Theta\right\}
		\end{equation}
		\STATE Let $k=0$ and choose $\tilde{\boldsymbol{p}}^{(0)}\in\left[\boldsymbol{0},\tilde{\boldsymbol{p}}_{\mathrm{max}}\right]$.
		\REPEAT
		\STATE Given $\tilde{\boldsymbol{p}}^{(0)}$, solve $\mu_j^{(k)}=\min\limits_{1\le i\le N_t}\frac{\tilde{f}_i\left(\tilde{\boldsymbol{p}}^{(k)}\right)}{\boldsymbol{z}_j(i)\hat{f}_i\left(\tilde{\boldsymbol{p}}^{(k)}\right)}$
		\STATE Given $\mu_j^{(k)}$, solve
		\begin{equation}\nonumber
			\tilde{\boldsymbol{p}}^{(k+1)}=\arg\max\limits_{\tilde{\boldsymbol{p}}\in\mathcal{F}}\min\limits_{1\le i\le N_t}\left(\tilde{f}_i(\tilde{\boldsymbol{p}})-\mu_j^{(k)}\boldsymbol{z}_j(i)\hat{f}_i(\tilde{\boldsymbol{p}})\right)
		\end{equation}
		\STATE $k=k+1$
		\UNTIL $\max\limits_{\tilde{\boldsymbol{p}}\in\mathcal{F}}\min\limits_{i}\left(\tilde{f}_i(\tilde{\boldsymbol{p}})-\mu_j^{(k)}\boldsymbol{z}_j(i)\hat{f}_i(\tilde{\boldsymbol{p}})\right)\le 0$
		\STATE The projection of $\boldsymbol{z}_j$ is given by $\boldsymbol{\pi}^{(j)}=\mu_j^{(k-1)}\boldsymbol{z}_j$
		\STATE $j=j+1$
		\UNTIL $\max\limits_i\left\{\left(\boldsymbol{z}_{j-1}(i)-\boldsymbol{\pi}_{\boldsymbol{z_{j-1}}}(i)\right)/\boldsymbol{z}_{j-1}(i)\right\}\le\delta_0$
		\STATE The optimal power allocation $\hat{\boldsymbol{p}}$ is obtained by solving $\pi^{(j-1)}_i=\frac{\tilde{f}_i(\hat{\boldsymbol{p}})}{\hat{f}_i(\hat{\boldsymbol{p}})}$
	\end{algorithmic}
\end{algorithm}

Since the objective function is an increasing function in $\boldsymbol{z}$, the optimal solution must occur at points where $\boldsymbol{z}_i=\frac{\tilde{f}_i(\tilde{\boldsymbol{p}})}{\hat{f}_i(\tilde{\boldsymbol{p}})}$. As a result, the optimal solution can be obtained by continually shrinking the region containing the feasible set and simultaneously finding the points that maximize the objective function within the region. When the region and the feasible set are sufficiently close, the optimal solution of the problem can be approximated by the optimal point within the region. In the MAPEL algorithm, the region is defined by its vertex set $\mathcal{V}$. At the beginning of the algorithm, the approximation factor $\delta_0$ is chosen decide the condition of termination is determined and the initial vertex is constructed by a vector $\boldsymbol{v^{(1)}}$ whose $i$th element is
\begin{equation}
	v_i^{(1)}=\max_{\tilde{\boldsymbol{p}}\in\left[\boldsymbol{0},\boldsymbol{\tilde{p}}_{\mathrm{max}}\right]}\frac{\tilde{f}_i(\tilde{\boldsymbol{p}})}{\hat{f}_i(\tilde{\boldsymbol{p}})}=1+\frac{\tilde{g}_i{{p}}_{\mathrm{max}}}{n},\forall i\in\left\{1,...,N_t\right\},
\end{equation} 
so that $\mathcal{Z}$ must be contained in the region $\left[\boldsymbol{0},\boldsymbol{v}^{(1)}\right]$. In each $j$th iteration, let $\mathcal{V}_j$ denote the current vertex set. Within this set, we denote by $\boldsymbol{z}^{(j)}$ the maximum point that also satisfies the minimum rate constraint, i.e. $\boldsymbol{z}^{(j)}=\arg\max\left\{\prod_{i=1}^{N_t}\left(v_i\right)^{\tilde{\omega}_i}|\boldsymbol{v}\in\mathcal{V}_j\cap\Theta\right\}$, where $\Theta=\left\{\boldsymbol{v}|v_i\ge 2^{{R}_{\mathrm{min}}}\right\}$ is the set of $\boldsymbol{z}$ that satisfies the rate constraint. The projection of $\boldsymbol{z}^{(j)}$ to the feasible region $\mathcal{Z}$ is computed to estimate the difference between the feasible set and the current region. Let $\boldsymbol{\pi}^{(j)}$ denote the projection of $\boldsymbol{z}^{(j)}$, the difference is given by
\begin{equation}
	\delta=\max\limits_i\left\{\left(z^{(j)}_i-\pi^{(j)}_i\right)/z^{(j)}_i\right\}.
\end{equation}
When $\delta$ is not greater than $\delta_0$, the difference is small enough and the algorithm terminates with $\boldsymbol{\pi}^{(j)}$ as an approximation of the optimal solution. Otherwise, a new vertex set $\mathcal{V}_{j+1}$ is constructed by replacing $\boldsymbol{z}^{(j)}$ with $N_t$ new vertices. Let $\boldsymbol{v}^{(j+1)}_m$ denote the $m$th new vertex, which is defined as
\begin{equation}
	\boldsymbol{v}^{(j+1)}_m=\boldsymbol{z}^{(j)}-\left(\boldsymbol{z}^{(j)}_m-\boldsymbol{\pi}^{(j)}_{m}\right)\boldsymbol{e}_m,
\end{equation}  
where $\boldsymbol{e}_m$ is the $m$th unit vector of $\mathcal{R}_+^{N_t}$. Because the upper boundary of $\mathcal{Z}$ is not explicitly known, the following max-min problem is formulated to calculate $\mu_j$ such that $\boldsymbol{\pi}^{(j)}=\mu_j\boldsymbol{z}^{(j)}$
\begin{equation}
	\begin{split}
		\mu_j&=\max\left\{c|c\boldsymbol{z}_j\in\mathcal{Z}\right\}\\
		&=\max\left\{c\le\min_{1\le i\le N_t}\frac{\tilde{f}_i(\tilde{\boldsymbol{p}})}{\hat{f}_i(\tilde{\boldsymbol{p}})},\tilde{\boldsymbol{p}}\in\mathcal{F}\right\}\\
		&=\max_{\tilde{\boldsymbol{p}}\in\mathcal{F}}\min_{1\le i\le N_t}\frac{\tilde{f}_i(\tilde{\boldsymbol{p}})}{\hat{f}_i(\tilde{\boldsymbol{p}})},
	\end{split}
\end{equation}
which is solved by modified Dinkelbach-type algorithm. 

When the algorithm converges, we denote by $\boldsymbol{\pi}^*$ the projection of the optimal $\boldsymbol{z}$. The optimal allocation $\boldsymbol{p}^*$ is obtained by solving the following equation:
\begin{equation}
	\boldsymbol{\pi}^*_i=\frac{\tilde{f}_i(\boldsymbol{p}^*)}{\hat{f}_i(\boldsymbol{p}^*)},
\end{equation}
and the average rate of the network is given by
\begin{equation}
	R_{\mathrm{ave}}=\sum_{i=1}^{N_t}\tilde{\omega}_{i}\log_2(\pi^*_i).
\end{equation} 
The detailed MAPEL algorithm is shown in Algorithm \ref{MAPEL}.
\subsection{Delay Constrained  massive Communication Networks}
In the last subsection, a general massive communication network is considered and its fundamental limit is obtained by mean-field approximation together with MAPEL algorithm. In this section, we focus on the massive networks with delay constraints. In particular, we consider a random data arrival process for each user and aim to investigate the delay constrained throughput of the network. We formulate the optimal power control problem as a MFG, which can be solved by primal dual hybrid gradient method effectively \cite{MFG}.

We denote time by $\tau$ and divided it into timeslots of duration $T$ seconds, which is equal to the delay constraint. At the beginning of each timeslot, let $a_i$ denote the number of bits arriving at the buffer of user $i$. We assume that the data arrival processes are i.i.d. across all users. The probability density function (p.d.f.) of $a_i$ is given by $f_A(a)$. At time $\tau$, we denote by $h_i(\tau)$ the and $g_{ji}(\tau)$ the channel gain between user $i$ and its destination node and the interference channel gain from user $j$ to user $i$, respectively. These gains are also assumed to be i.i.d. across all users, with pdfs $f_H(h)$ and $f_G(g)$, respectively. Let $p_i(\tau)$  denote the power allocation to user $i$ at time $\tau$. The maximum transmission rate of user $i$ is given by
\begin{equation}
	r_i(\tau)=\log_2\left(1+\frac{p_i(\tau)h_i(\tau)}{\sum_{j\ne i}{p_j(\tau)g_{ji}(\tau)+n}}\right),
\end{equation}  
Let $s_i(\tau)$ denote the number of bits in the buffer of user $i$. The buffer dynamics are given by
\begin{equation}
	ds_i=-r_id\tau.
\end{equation}

We aim to minimize the power to clear the buffer of all users while satisfying the delay constraint. As a result, the following power control problem is formulated
\begin{subequations}
	\label{delay}
	\begin{align}
		\min_{p_i(\tau)}\quad\quad\sum_{i=1}^{N_o}\int_{0}^{T}&p_i(\tau)d\tau \\
		s.t. \quad r_i(\tau)=\log_2&\left(1+\frac{p_i(\tau)h_i(\tau)}{\sum_{j\ne i}p_j(\tau)g_{ji}(\tau)+n}\right),\\
		ds_i=&-r_id\tau,\\
		s_i(0)&=a_i,\\
		s_i(T)&=0,
	\end{align}
\end{subequations}
where $N_o$ is the total number of users. However, in massive communication scenarios, large number of users results in very high computational complexity when solving the problem directly. To address this, we will to reformulate Problem (\ref{delay}) as a MFG, which focuses on the average behavior of the users instead of the power allocation of each user separately. 

To construct the MFG problem, we define $\rho(s,h,\tau)$ as the joint p.d.f. of the number of bit in the buffer and the channel state at time $\tau$. It satisfies
\begin{equation}
	\int_{h_1}^{h_2}\int_{s_1}^{s_2}\rho(s,h,\tau_0)=\mathrm{Pr}\left\{s(\tau_0)\in(s_1,s_2),h(\tau_0)\in(h_1,h_2)\right\}.
\end{equation}
Let $p(s,h,\tau)$ denote the power allocated to the links with buffer containing $s$ bits and channel gain $h$ at time $\tau$. Then the transmission rate of these links is given by
\begin{equation}\label{rate}
	\begin{split}
		r(s,h,\tau)&=\log_2\left(1+\frac{
			p(s,h,\tau)h}{N_o\int_{h}\int_{s}\int_{g}gpf_G(g)\rho dgdsdh+n}\right),\\
		&=\log_2\left(1+\frac{p(s,h,\tau)h}{\bar{g}\int_{h}\int_{s}p\rho dsdh+n}\right),
	\end{split}
\end{equation}
where $\bar{g}=N_o\int_{g}gf_G(g)dg$ is the average gain of interference channel. Besides, we assume that the channel gains experience fast fading which is given by
\begin{equation}
	dh(\tau)=\eta dW(\tau),
\end{equation}
where $W(\tau)$ is the standard Wiener process and $eta$ is the coefficient to represent the fading speed. Combining the dynamics of data transmission (which affect the buffer state $s$) with those of channel fading (which affect the channel gain $h$), the evolution of the distribution $\rho(s,h,\tau)$ of the users is described by the following partial differential equation:
\begin{equation}
	\frac{\partial\rho}{\partial \tau}-\frac{\partial(\rho r)}{\partial s}-\eta\frac{\partial^2\rho}{\partial h^2}=0.
\end{equation}
Meanwhile, the total power in  (\ref{delay}a) is given by
\begin{equation}\label{power}
	P_{total}=\int_{0}^{T}\int_{h}\int_{s}p(s,h,\tau)\rho(s,h,\tau)dsdhd\tau.
\end{equation}
The complete MFG for delay constrained network in then presented in the following theorem.
\begin{theorem}
	The MFG for Problem (\ref{delay}) is given by
	\begin{subequations}\label{MFG_p}
		\begin{align}
			\min_{p,\rho}\int_{0}^{T}&\int_{h}\int_{s}p\rho dsdhd\tau\\
			s.t. \quad \frac{\partial\rho}{\partial \tau}&-\frac{\partial(\rho r)}{\partial s}-\eta\frac{\partial^2\rho}{\partial h^2}=0\\
			r=\log_2\Big(1&+\frac{p(s,h,\tau)h}{\bar{g}\int_{h}\int_{s}p\rho dsdh+n}\Big)\\
			\int_h&\rho(0,h,T)dh=1\\
			&\rho(s,h,0)=\rho^0			
		\end{align}
	\end{subequations}
\end{theorem}

In order to solve the above problem, we define the Lagrange Multiplier $\Phi(s,h,\tau)$. Then the Lagrange Dual problem of  Problem (\ref{MFG_p}) is given by
\begin{equation}\label{L}
	\inf_{p,\rho}\sup_{\Phi}L(p,\rho,\Phi)
\end{equation}
where the Lagrangian $L$ is
\begin{equation}
		L(p,\rho,\Phi)=\int_{0}^{T}\int_{h}\int_{s}\left (p\rho-\Phi\left(\frac{\partial\rho}{\partial \tau}-\frac{\partial(\rho r)}{\partial s}-\eta\frac{\partial^2\rho}{\partial h^2}\right)\right)dsdhd\tau.
\end{equation}
This Lagrangian dual problem can then be solved by updating $\rho(s,h,\tau)$, $r(a,h,\tau)$ and $\Phi(s,h,\tau)$ as follows
\begin{subequations}
	\begin{align}
		\nonumber p^{k+1}&=\arg\min_{p}L_1\left(p,\rho,\Phi^k,p^k\right) \\ 	&=\arg\min_{p}\left\{L(p,\rho,\Phi^k,p^k)+\frac{1}{2\varrho}\left\|p-p^k\right\|_2^2\right\},\\
		\nonumber \rho^{k+1}&=\arg\min_{\rho}L_2\left(r^{k+1},\rho,\Phi^k,\rho^k\right)\\ 
		&=\arg\min_{\rho}\left\{L(r^{k+1},\rho,\Phi^k)+\frac{1}{2\varrho}\left\|\rho-\rho^k\right\|_2^2\right\},\\
		\nonumber \Phi^{k+1}&=\arg\max_{\Phi}L_3\left(2r^{k+1}-r^k,2\rho^{k+1}-\rho^k,\Phi,\Phi^k\right)\\
		\nonumber&=\arg\max_{\Phi}\bigg\{L\left(2r^{k+1}-r^k,2\rho^{k+1}-\rho^k,\Phi\right)\\	
		&-\frac{1}{2\varsigma}\left\|\Phi-\Phi^k\right\|^2_{H_1}\bigg\},
	\end{align}
\end{subequations}
where $\varrho$ and $\varsigma$ are the step sizes and the $H_1$ norm is defined as
\begin{equation}
	\left\|f\right\|^2_{H_1}=\int_{0}^{T}\int_{h}\int_{s}\left(\left(\partial_{\tau}f(s,h,\tau)\right)^2+\left\|\nabla f(s,h,\tau)\right\|^2\right)dsdhd\tau.
\end{equation}
The above iteration can be solved by differentiating the equations with $p$, $\rho$ and $\Phi$, respectively. The optimal solution of Problem (\ref{MFG_p}) can be then obtained by repeating the iteration until convergence. 

\section{The capacity of infrastructure-free networks}
\label{IF}
In this section, we first provide a simple but efficient routing strategy for infrastructure-free networks. Based on the routing strategy, we prove that the average rate of the infrastructure-free network is proportional to the transport capacity of the corresponding single-hop network. The optimal power allocation and the capacity of the infrastructure-free network are then derived by exploring the optimal parameter of the routing strategy and utilizing the results in Section \ref{SH}.
\subsection{Efficient Routing Strategy}
In this subsection, we present a simple yet efficient routing strategy named optimal equidistant path  approximation (OEPA) strategy  to decompose a multi-hop network into a single-hop network. In a multi-hop Scheme, nodes in the network can operate as relay nodes,  forwarding data from source to destination. When selecting these relay nodes, the following conditions must be satisfied. \begin{enumerate} \item  The relay nodes should be as close as possible to the line joining $T_i$ and $R_i$. \item The distances between the neighboring relay nodes should be as uniform as possible. \end{enumerate}The first condition is to reduce the total distance from source node to destination, which is denoted by $\tilde{d}_{ii}$, via the relay nodes. When all relay nodes are on the line joining $T_i$ and $R_i$, this distance is minimized and equal to the distance between $T_i$ and $R_i$, i.e. $\tilde{d}_{ii}=d_{ii}$. The second condition ensures that the rates between relay nodes are as uniform as possible, thereby improving the power efficiency of the system. This is grounded in the fact that the path loss between the two nodes is determined by the distance between them, which the capacity of the link is determined by its bottleneck. When the relay nodes follows a nearly uniform distribution, the power attenuation between the neighbor nodes is approximately equal, assuming the small scale fading is neglected. Consequently, and the power can be allocated evenly across different segments, leading to an equal rate in each segment of the link. In this case, the power efficiency is maximized and the network can achieve a higher rate under the same power constraint.  In the following, we aim to present the OEPA strategy that satisfies the above conditions. 

In the OEPA strategy, let $T_i$ and $R_i$ denote the source node and destination node, respectively, for the $i$th link. Let $N_{i}^r$ denote the number of relay nodes for this link. When the expected distance between the neighboring relay nodes is $r_0$, the number of relay nodes is given by  $N_{i}^r=\left\lceil \frac{\left\|R_i-T_i\right\|}{r_0}\right\rceil-1$, where $\left\lceil a\right\rceil$ is the minimum integer greater than $a$.  We denote by $\tilde{X}_j^i$ the desirable locations of the $j$th relay node along the path from the source node $T_i$ to the destination node $R_i$, given by $\tilde{X}_j^i=T_i+jr_0\frac{(R_i-T_i)}{\left\|R_i-T_i\right\|}$. The relay nodes are then chosen as the nearest nodes to each desirable locations. Let $X_j^i$ denote the location of the $j$th relay node. We have 
\begin{equation}
	X_{j}^i = \arg\min_{X\in\mathcal{N}}\left\|X-\tilde{X}_j\right\|,
\end{equation}
where $\mathcal{N}$ is the set of all nodes in the network. In practical networks, the distance between $T_i$ and $R_i$ may not be an integer multiple of $r_0$. In this case, the distance between the last relay node and the destination node is reduced to $\left\|R_i-T_i\right\|-N_{i}^rr_0$. Meanwhile, let $N_{i}^h=N_{i}^r+1$ denote the number of hops for the $i$th link. The probability that it takes $a$ hops from $T_i$ to $R_i$ is given by
\begin{equation}\label{N_ll}
	\begin{split}
		\mathrm{Pr}\left\{N^{h}_i=a\right\}&=\mathrm{Pr}\left\{(a-1)r_0<d_{ii}\le ar_0\right\}\\
		&=\frac{(2a-1)r_0^2}{d_0^2{N_m}^2}.
	\end{split}
\end{equation} 
Since the destination nodes of the links are chosen randomly, $N_{i}^h$ are i.i.d. across $\forall i\in\left\{1,...,N_o\right\}$. The maximum number of $N_{h}$ is given by $N^{h}_{\mathrm{max}}=\left\lceil \frac{d_0{N_m}}{r_0}\right\rceil$.  

\subsection{Equivalent Single-Hop Network}
Under the OEPA strategy, the original network is transformed to a single-hop network, where the distance of each hop is approximately equal to $r_0$. However, since the relay nodes are distributed randomly, the actual locations of the relay nodes will deviate the desirable locations. At a result, the distance of each hop is a random variable rather than a constant. To investigate the distribution of single hop distance, we present the distribution of deviation distance between the actual locations and the desirable locations of the relay nodes in the following lemma.
\begin{lemma}
	\label{lemma3}
	The distance between the desirable locations and the actual locations og the relay nodes follows a Rayleigh distribution with parameter $\sigma_d^2=\frac{1}{2\pi\lambda}$. 
\end{lemma}
\begin{IEEEproof}
	Let $\hat{d}$ denote the distance between desirable relay location the real location, which is given by $\hat{d}=\left\|\tilde{X}-X\right\|$. The probability that $\hat{d}$ is greater than $a$ is the same as the probability that there is no node within the circle of radius $a$ centered at $\tilde{X}$, which is given by
	\begin{equation}\label{PPP}
		\text{Pr}\left\{\hat{d}>a\right\}=e^{-\lambda\pi a^2}.
	\end{equation}     
	(\ref{PPP}) is based on the fact that the nodes are distributed according to Poisson point process. Therefore, $\text{Pr}\left\{\hat{d}<a\right\}=F_{\hat{d}}(a)=1-e^{-\lambda\pi a^2}$. The pdf of $\hat{d}$ is given by
	\begin{equation}
		f_{\hat{d}}(a)=\frac{\mathrm{d}F_{\hat{d}}(a)}{\mathrm{d} a}=e^{-\lambda\pi a^2}2\pi\lambda a,
	\end{equation}
	which is Rayleigh distribution with parameter $\sigma_d^2=\frac{1}{2\pi\lambda}$. This completes the proof.
\end{IEEEproof}
Let $\hat{d}_x$ and $\hat{d}_y$ denote the horizontal and vertical components of $\hat{d}$, which satisfy ${\hat{d}}^2=\hat{d}_x^2+\hat{d}_y^2$. Since the nodes are distributed randomly, $\hat{d}_x$ and $\hat{d}_y$ are i.i.d.. As a result, $\hat{d}_x$ and $\hat{d}_y$ are Gaussian random variables with zero mean and $\sigma_d^2$ variance, i.e. $\hat{d}_x, \hat{d}_y\sim\mathcal{N}\left(0,\frac{1}{2\pi\lambda}\right)$. 

Let $d_{sr}, d_{rr}$ and $d_{rd}$ denote the distances between the source node and first relay node, the distance between neighboring relay nodes, and the distance between the last relay node and destination node, respectively. We consider an ideal situation where the distance between the source node and the destination node is an integer multiple of $r_0$. In this case, $d_{sr}$ and $d_{rd}$ are identically distributed and we denote them by $d_{nr}$. The distribution of $d_{rr}$ and $d_{nr}$ in case of the large the nodes intensity is presented in the following lemma.
\begin{lemma}
	\label{lemma4}
	When the intensity of nodes is large, the distance between the relay node and source node or destination node follows a Gaussian distribution with mean $r$ and variance $\frac{1}{2\pi\lambda}$. The distance between the neighbor relay nodes follows a Gaussian distribution with mean $r$ and variance $\frac{1}{\pi\lambda}$, i.e., $d_{nr}\sim\mathcal{N}\left(r,\frac{1}{2\pi\lambda}\right)$ and $d_{rr}\sim\mathcal{N}\left(r,\frac{1}{\pi\lambda}\right)$.
\end{lemma}
\begin{IEEEproof} 
	The two distances are given by
	\begin{equation}
		\begin{split}
			d_{nr}&=\sqrt{\left(r_0+\hat{d}_x\right)^2+\hat{d}_y^2}\\
			&=r_0\left({1+\frac{2\hat{d}_x}{r_0}+\left(\frac{\hat{d}_x}{r_0}\right)^2+\left(\frac{\hat{d}_y}{r_0}\right)^2}\right)^{\frac{1}{2}},
		\end{split}
	\end{equation}  
	and
	\begin{equation}
		\begin{split}
			&d_{rr}=\sqrt{\left(r_0+\hat{d}_{1x}-\hat{d}_{2x}\right)^2+\left(\hat{d}_{1y}-\hat{d}_{2y}\right)^2}\\
			&=r_0\sqrt{(1+\frac{2\left(\hat{d}_{1x}-\hat{d}_{2x}\right)}{r_0}+\left(\frac{\hat{d}_{1x}-\hat{d}_{2x}}{r_0}\right)^2+\left(\frac{\hat{d}_{1y}-\hat{d}_{2y}}{r_0}\right)^2},
		\end{split}
	\end{equation}  
	where $\hat{d}_{1}$ and $\hat{d}_{2}$ are the deviation distances of the two relay nodes, respectively. When the intensity of nodes for the network is large, the deviation is much smaller than the desirable single-hop distance between, i.e., $\hat{d}<<r_0$. Therefore, the higher order term in above equations can be ignored. We have
	\begin{equation}
		\begin{split}
			d_{nr}&=r_0\sqrt{1+\frac{2\hat{d}_x}{r_0}+\left(\frac{\hat{d}_x}{r_0}\right)^2+\left(\frac{\hat{d}_y}{r_0}\right)^2}\\
			&\approx r_0\sqrt{1+\frac{2\hat{d}_x}{r_0}}\\
			&\approx r_0+\hat{d}_x,
		\end{split}
	\end{equation}
	and
	\begin{equation}
		\begin{split}
			d_{rr}&\approx r_0\sqrt{1+\frac{2\left(\hat{d}_{1x}-\hat{d}_{2x}\right)}{r_0}}\\
			&\approx r_0+(\hat{d}_{1x}-\hat{d}_{2x}).
		\end{split}
	\end{equation}
	The proof is completed.
\end{IEEEproof}
Although the distances of the hops for this ideal equivalent single hop (IESH) network are not strictly independent. Since only the distances of the hops sharing a common relay node are correlated, and the total number of hops are sufficiently large for large-scale networks, the distance for each hop can be considered to be independent. Let $\eta_n$ and $\eta_r$ denote the portion of distances $d_{nr}$ and $d_{rr}$ in the IESH network. For each link that consists of $N_h$ hops, since the distances of first hop and last hop are given by $d_{nr}$ while the distances of the rest hops are given by $d_{rr}$, $\eta_n$ is given by 
\begin{equation}\label{portion}
	\begin{split}
		\eta_n&=\mathrm{Pr}\left\{N^h=1\right\}+\sum_{i=2}^{N^h_{\mathrm{max}}}\frac{2}{i}\mathrm{Pr}\left\{N^h=i\right\}\\
		&=\frac{r_0^2}{d_0^2{N^c_m}^2}+\sum_{i=2}^{N^h_{\mathrm{max}}}\frac{2}{i}\frac{(2i-1)r_0^2}{d_0^2{N_m}^2}.
	\end{split}
\end{equation}
Besides, $\eta_r$ is given by $\eta_r=1-\eta_s$.
In the following, we derive the distribution of single-hop distance for IESH network. We first quantize $d_{nr}$ and $d_{rr}$ to a common set which is denoted as $\left\{d^s_{1}, d^s_{2},..., d^s_{N_{ds}}\right\}$ where $N_{ds}$ is the total number of elements in the set. Let $\varepsilon^{rr}_i$ and $\varepsilon^{nr}_i$ denote the probabilities that $\mathrm{Pr}\left\{d_{nr}=d_{si}\right\}$ and $\mathrm{Pr}\left\{d_{rr}=d_{si}\right\}$. We denote by $f_{a,b}(x)$ the pdf of Gaussian distribution with mean $a$ and variance $b$. The two probabilities are given by
\begin{equation}
	\varepsilon^{rr}_i=\int_{\frac{d^s_{i-1}+d^s_i}{2}}^{\frac{d^s_i+d^s_{i+1}}{2}}f_{r,\frac{1}{\pi\lambda}}(x)\mathrm{d}x,
\end{equation}
and
\begin{equation}
	\varepsilon^{nr}_i=\int_{\frac{d^s_{i-1}+d^s_i}{2}}^{\frac{d^s_i+d^s_{i+1}}{2}}f_{r,\frac{1}{2\pi\lambda}}(x)\mathrm{d}x,
\end{equation}
where $d^s_0=-\infty$ and $d^s_{N_{d_s}+1}=\infty$. Let $d_s\in\left\{d^s_{1}, d^s_{2},..., d^s_{N_{ds}}\right\}$ denote the single-hop distance for IESH network. We denote by $\varepsilon^h_i$ the probability that $\mathrm{Pr}\left\{d_s=d^s_i\right\}$, which is given by
\begin{equation}
	\varepsilon^r_i=\eta_n\varepsilon^{nr}_i+\eta_r\varepsilon^{rr}_i.
\end{equation}

Under the multi-hop scheme, the rate of the $i$th link is defined as the ratio of the number of transmitted bits to the average required time. Meanwhile, the rate of the network is defined as the average rate of all links. Let $\bar{r}_m$ and $\bar{r}_I$ denote the average of rate of the infrastructure-free network and the corresponding IESH network, respectively. We present the relationship between them in the following theorem.
\begin{theorem}
	\label{theorem2}
	Let $\bar{r}_I$ and $r_0$ denote the average rate and the desirable single-hop distance of IESH network, respectively. The average rate of the infrastructure-free network can be approximated as  
	\begin{equation}\label{Theorem 2}
		\bar{r}_m\approx \frac{2r_0\bar{r}_I}{d_0^2{N_m}}.
	\end{equation}
\end{theorem}
\begin{IEEEproof}
	We first consider a new multi-hop network which is almost identical with the original one. The only difference is that, for the new network, the distances between the source and the destination nodes  are integer multiple of $r_0$ with distribution
	\begin{equation}\label{IESH_u}
		\mathrm{Pr}\left\{d_{ii}^u=ar_0\right\}=
		\frac{(2a+1)r_0^2}{d_0^2{N_m}^2},1\le a<N_h^u,
	\end{equation} 
	where $N_h^u=\left\lfloor \frac{{N_m}d_0}{r}\right\rfloor$. By applying the OEPA strategy, the new network is transformed to a single-hop network which is noted as the IESH-u network. Comparing  (\ref{N_ll}) and  \ref{IESH_u}), we have
	\begin{equation}
		\mathrm{Pr}\left\{d_{ii}^u=ar_0\right\}>\mathrm{Pr}\left\{N_h=a\right\},
	\end{equation}
	for $1\le a\le\left\lfloor \frac{{N_m}d_0}{r_0}\right\rfloor$. This indicates that the number of hops for the IESH-u network is smaller than the original network. According to  (\ref{portion}), we have $\eta_n^u>\eta_n$ and $\sigma_{d_s^u}^2<\sigma_{d_s}^2$, where $\eta_n^u$ is the portion of $d_{nr}$ for the IESH-u network, $\sigma_{d_s^u}^2$ and $\sigma_{d_s}^2$ are the variance of single-hop distance of the IESH-u network and the IESH network, respectively. As a result, the average rate of the IESH-u network is greater than the average rate of the IESH network, i.e. $\bar{r}^u_I>\bar{r}_I$. Meanwhile, since the number hops of the new network is smaller than the original network, the average rate of the new network is greater than the original network. Let $\bar{r}_m^u$ denote the average rate of the new network, we have
	\begin{equation}\label{upper}
		\begin{split}
			\bar{r}_m^u&=\sum_{i=1}^{N_h^u}\frac{\bar{r}^u_I}{i}\mathrm{Pr}\left\{d_{ii}^u=ar_0\right\}\\
			&>\sum_{i=1}^{N_h^u}\frac{\bar{r}^u_I}{i}\frac{(2i+1)r_0^2}{d_0^2{N_m}^2}\\
			&>\bar{r}_m.
		\end{split}
	\end{equation}	
	Meanwhile, we consider another multi-hop network, which is also almost identical with the original network. The only difference is that, for this multi-hop network, the distances between the source and the destination nodes are integer multiple of $r_0$ with distribution
	\begin{equation}
		\begin{split}
			\mathrm{Pr}\left\{d_{ii}^l=ar_0\right\}&=\mathrm{Pr}\left\{N^h=a\right\}\\
			&=\frac{(2a-1)r_0^2}{{N_m}^2d_0^2},1\le a<N_h^l
		\end{split}
	\end{equation} 
	where $N_h^l=N^h_\mathrm{max}=\left\lceil\frac{d_0{N_m}}{r_0}\right\rceil$. By applying the OEPA strategy, this network is transformed to a single-hop network, which is noted as the IESH-l network. It is obvious that the IESH-l network is the same as the IESH network.  Hence, we have $\bar{r}^l_I=\bar{r}_I$, where $\bar{r}^l_I$ is the average rate of the IESH-l network. However, because the distance of the last hop of the original network is smaller than that of the new network, the average rate of the original network is greater than that of this new network. Let $\bar{r}_m^l$ denote the average rate of the new network, we have
	\begin{equation}\label{lower}
		\begin{split}
			\bar{r}_m&>\bar{r}_m^l\\
			&=\sum_{i=1}^{N_h^l}\frac{\bar{r}_I}{i}\mathrm{Pr}\left\{d_{ii}^l=ar_0\right\}\\
			&=\sum_{i=1}^{N_h^l}\frac{\bar{r}_I}{i}\frac{(2i-1)r_0^2}{d_0^2{N_m}^2}.
		\end{split}
	\end{equation}
	Combining  (\ref{upper}) and (\ref{lower}), we have
	\begin{equation}	\frac{2r_0\bar{r}_I}{d_0^2{N_m}}-\sum_{i=1}^{N_{h}^{l}}\frac{1}{i}\frac{r_0^2\bar{r}_I}{d_0^2N_m^2}\le \bar{r}_m\le\frac{2r_0\bar{r}_I}{d_0^2{N_m}}+\sum_{i=1}^{N_{h}^{u}}\frac{1}{i}\frac{r_0^2\bar{r}_I}{d_0^2N_m^2}.
	\end{equation}
	Let $\Xi=\frac{2r\bar{r}_I}{d_0^2{N_m}}$,  we can rewrite the above equation as 
	\begin{equation}
		1-\sum_{i=1}^{N_{h}^{l}}\frac{1}{i}\frac{r_0}{2{N_m}}\le\frac{\bar{r}_m}{\Xi}\le1+\sum_{i=1}^{N_{h}^{u}}\frac{1}{i}\frac{r_0}{2{N_m}}
	\end{equation}
	In case that ${N_m}$ is sufficiently large, the summation part of the above equation converges to 0, and we have
	\begin{equation}
		\frac{\bar{r}_m}{\Xi}\approx1.
	\end{equation} 
	This completes the proof.
\end{IEEEproof}
Theorem \ref{theorem2} indicates that the average rate of the infrastructure-free network is proportional to the product of the rate of the corresponding IESH network and its single-hop distance. This product is defined as the network transport capacity in \cite{capacity_gupta}. In the next subsection, we aim to investigate the optimal power allocation to maximize 
the transport capacity of the IESH network. 
\subsection{IESH-s network}
In this subsection, we consider a simple case of the IESH networks where the small fading factor is ignored, i.e. $G(d)=f(d)$, which is named as IESH-s network. A more general case where the channel gains suffer small scale fading is discussed in the next subsection. In the IESH-s network, the channel gain between the nodes is fixed and the achievable rate of the network is limited by the minimum rate. Therefore, the power is allocated so that the rate between all neighboring nodes is equal. We aim to derive the optimal single-hop distance and power allocation strategy to maximize the transport capacity of the the IESH-s networks.  As a result, the following optimization problem is formulated.
\begin{algorithm}[t]
\caption{Algorithm for transport capacity for IESH-s network}
\label{IESH-s}
\renewcommand{\algorithmicrequire}{\textbf{Input:}}
\renewcommand{\algorithmicensure}{\textbf{Output:}}
\begin{algorithmic}[1]
	\STATE \textbf{Initialization} Let $i=1$, $\tau=0.618$ and Choose the minimum single-hop distance $r^{\mathrm{min}}_i$ and the maximum single-hop distance $r^{\mathrm{max}}_i$.
	\REPEAT
	\STATE $r^l_i=r^{\mathrm{min}}_i+(1-\tau)\left(r^{\mathrm{max}}_i-r^{\mathrm{min}}_i\right)$ and $r^u_i=r^{\mathrm{min}}_i+\tau\left(r^{\mathrm{max}}_i-r^{\mathrm{min}}_i\right)$
	\STATE Find the maximum information rate of the IESH-s network with single-hop distance $r^l_i$ and $r^u_i$ by bisection method according  (\ref{F_matrix}) and (\ref{feasibility}) and denote them by $R^l_i$ and $R^u_i$, respectively. 
	\IF{$r^l_iR^l_i>r^u_iR^u_i$}
	\STATE Let $r^{\mathrm{min}}_{i+1}=r^{\mathrm{min}}_{i}$ and $r^{\mathrm{max}}_{i+1}=r^u_i$.
	\ELSE
	\STATE Let $r^{\mathrm{min}}_{i+1}=r^u_i$ and $r^{\mathrm{max}}_{i+1}=r^{\mathrm{max}}_{i}$
	\ENDIF
	\STATE $i=i+1$
	\UNTIL{$r^{\mathrm{max}}_i-r^{\mathrm{min}}_i\le \delta_r$}
\end{algorithmic}
\end{algorithm}
\begin{subequations}\label{IESH_s_ori}
\begin{align}
	\max_{r_0,\boldsymbol{p}}\quad &r_0r\\
	s.t.\ \log_2\Big(1+&\frac{p_iG_{ii}}{\sum_{j\ne i}p_jG_{ji}+n}\Big)\ge r,\\ 
	\frac{1}{N_o}\sum_{i=1}^{N_s}p_i&\le p_{\mathrm{ave}}.
\end{align}
\end{subequations}
for $\forall i\in\left\{1,...,N_o\right\}$. In the IESH-s network, since the channel gain between the nodes is fixed, we consider the average power constraint. In the following, we derive the mean-fied approximation of Problem (\ref{IESH_s_ori}).

Let $G_{ii}\in\left\{g_1,...g_{N_{ms}}\right\}$ denote the channel gain for each link in the IESH-s network. Since the small scale fading is ignored, the channel gain is determined by the distance between the nodes. Thereby, we have $g_i=f(d_i)$ and $N_{ms}=N_{ds}$. Meanwhile, the probability that $\mathrm{Pr}\left\{G_{ii}=g_i\right\}$ is equal to $\varepsilon^r_i$. Let $G_{ij}\in\left\{g^{I}_1,...,g^{I}_{N^I_{gs}}\right\}, j\ne i$ denote the interference channel gain, where $g^{I}_i=f(id_0)$. The probability that $\mathrm{Pr}\left\{G_{ij}=g^{I}_i\right\}$ is denoted as $\theta_i$ which is equal to $\gamma_i^I$ in (\ref{gamma_a_I}). Given the channel gain $\boldsymbol{g}$ and the interference channel gain $\boldsymbol{g}^{I}$, similar operations are applied to formulate the mean-field approximation of Problem (\ref{IESH_s_ori}). Let $\tilde{\boldsymbol{G}}$, $\boldsymbol{\tilde{g}}$ and $\boldsymbol{\tilde{\omega}}$ denote the equivalent interference channel gain matrix, channel gain vector and weight vector of the IESH-s network,
Problem (\ref{IESH_s_ori}) can be reformulated as 
\begin{subequations}\label{IESH_s_mf}
\begin{align}
	\max_{r_0,\boldsymbol{p}}\quad &r_0\tilde{r}\\
	s.t.\ \log_{2}\Big(1+&\frac{p_i\tilde{g}_i}{\sum_{j}p_j\tilde{G}_{ji}+n}\Big)\ge \tilde{r},\\
	\sum_{i}\tilde{\omega}_i&p_i\le p_{\mathrm{ave}}.
\end{align}
\end{subequations}
In the following, we aim to propose an alternating optimization algorithm to solve Problem  (\ref{IESH_s_mf}). 

In Problem (\ref{IESH_s_mf}), for the given expected single-hop distance, the maximum network rate is obtained by increasing the network rate until the rate or the required average power is infeasible. The feasibility check matrix is given by  (\ref{F_matrix}) while the corresponding power allocation vector is given by  (\ref{feasibility}). Since the transport capacity is concave with the expected single-hop distance, the optimal $r_0$ can be obtained by Golden section search method. The detailed algorithm to solve Problem (\ref{IESH_s_mf}) is presented in Algorithm \ref{IESH-s}. 

Based on Algorithm \ref{IESH-s}, we have obtained the transport capacity for the IESH-s network. The capacity for multi-hop network with fixed channel gain can be obtained by Theorem \ref{theorem2}. 

\subsection{IESH-g network} 
In this subsection, we consider a more general case of the IESH network, where the channel gain between the $T_i$ and $R_j$ is given by $G_{ij}=h_{ij}g(d_{ij})$. We note this network as the IESH-g network. In the IESH-g network, the channel between the nodes is not fixed due to the small scale fading. Therefore, the power is allocated to maximize the average rate of the network. As a result, the following optimization problem is formulated. 
\begin{subequations}\label{IESH_g_ori}
\begin{align}
	\max_{r_0,\boldsymbol{p}}\ \frac{r_0}{N_l}\sum_{i=1}^{N_l}\log_{2}\Bigg(&1+\frac{p_{i}G_{ii}}{\sum_{j\ne i }G_{ji}p_j+n}\Bigg)\\
	s.t.\quad \log_{2}\Bigg(1+&\frac{p_{i}G_{ii}}{\sum_{j\ne i }G_{ji}p_j+n}\Bigg)\ge R_{\mathrm{min}},\\
	\quad p_i&\le p_{\mathrm{max}}.
\end{align}
\end{subequations}
\begin{algorithm}[t]
\caption{Algorithm for transport capacity for IESH-s network}
\label{IESH_g}
\renewcommand{\algorithmicrequire}{\textbf{Input:}}
\renewcommand{\algorithmicensure}{\textbf{Output:}}
\begin{algorithmic}[1]
	\STATE \textbf{Initialization} Let $i=1$, $\tau=0.618$ and Choose the minimum single-hop distance $r^{\mathrm{min}}_i$ and the maximum single-hop distance $r^{\mathrm{max}}_i$.
	\REPEAT
	\STATE $r^l_i=r^{\mathrm{min}}_i+(1-\tau)\left(r^{\mathrm{max}}_i-r^{\mathrm{min}}_i\right)$ and $r^u_i=r^{\mathrm{min}}_i+\tau\left(r^{\mathrm{max}}_i-r^{\mathrm{min}}_i\right)$
	\STATE Find the maximum information rate of the IESH-g network with single-hop distance $r^l_i$ and $r^u_i$ by MAPEL algorithm and denote them by $R^l_i$ and $R^u_i$, respectively. 
	\IF{$r^l_iR^l_i>r^u_iR^u_i$}
	\STATE Let $r^{\mathrm{min}}_{i+1}=r^{\mathrm{min}}_{i}$ and $r^{\mathrm{max}}_{i+1}=r^u_i$.
	\ELSE
	\STATE Let $r^{\mathrm{min}}_{i+1}=r^u_i$ and $r^{\mathrm{max}}_{i+1}=r^{\mathrm{max}}_{i}$
	\ENDIF
	\STATE $i=i+1$
	\UNTIL{$r^{\mathrm{max}}_i-r^{\mathrm{min}}_i\le \delta_r$}
\end{algorithmic}
\end{algorithm}
To solve Problem (\ref{IESH_g_ori}), we need to obtain the channel gains between the nodes and the distribution of them. The small scale fading factor and its distribution are given by $h\in\left\{h_1, h_2,...,h_B\right\}$ and $\mathrm{Pr}\left\{h=h_b\right\}=\beta_b$, respectively. Meanwhile, the distances between the nodes and their distribution are given by $d_s=\left\{d^s_1,..., d^s_{N_{ds}}\right\}$ and $\mathrm{Pr}\left\{d_s=d^s_i\right\}=\varepsilon^r_i$. Similar to that in the single-hop network, let $\boldsymbol{g}=\left(g_{11}, g_{12},...,g_{ij},...,g_{N_hN_{ds}}\right)$ denote the channel gain between the nodes, where $g_{ij}=h_ig(d^s_j)$. We denote by $\theta_{ij}$ the probability that $\mathrm{Pr}\left\{g=g_{ij}\right\}$ which is given by $\theta_{ij}=\beta_i\varepsilon^r_j$. For simplicity, we reorder $g_{ij}$ in descending order and relabel them with a new subscript $k$ which is given by $k=(j-1)B+i$ for $k\in\left\{1,...,N_{g}\right\}, N_{g}=N_hN_{ds}$. Meanwhile, the probability $\theta_{ij}$ is relabeled with the same subscript of $g_{ij}$. The interference gain for the IESH-g network is the same as that in the single-hop network which is given by $g^{I}=\left\{g^I_1,..., g^I_{N_g^I}\right\}$. Similar to that in the single-hop network, we use $\boldsymbol{a}$ to describe the interference for each node and categorize them into $N_I$ groups, where the interference power for the nodes in the $i$th category is denoted as $I_i$ which is given by (\ref{I1}) and (\ref{I2}). Given the channel gain and the categories of interference power, we categorize all nodes into $N_t=N_{gm}N_I$ groups. The detailed process is similar to that in the single-hop network and omitted here. The channel gain and interference power in the $k$th group is given by $\tilde{g}_i$ and $\tilde{I}_i$, respectively. The probability that the node is in the $k$th group is given by $\tilde{\omega}_i$. The mean-field approximation of Problem (\ref{IESH_g_ori}) is given by 
\begin{subequations}\label{IESH_g_MF}
\begin{align}
	\max_{\tilde{\boldsymbol{p}},r_0}\quad&\sum_{i=1}^{N_t}r_0\tilde{\omega}_{i}\log_{2}\left(1+\frac{\tilde{p}_i\tilde{g}_i}{\tilde{I}_i+n}\right)\\\nonumber
	s.t.\quad \ &\tilde{I}_i =\sum_{j=1}^{N_a}\sum_{k=1}^{N_t}\bar{a}_{j\tilde{u}^i_j}\tilde{Q}^1_{jk}\tilde{p}_kg^{I}_j\\
	& \ \ +\sum_{j=N_a+1}^{N_g^I}\sum_{k=1}^{N_t}\tilde{N}_r^i\theta^I_kq_k\tilde{p}_kg_j^{I}, \\
	&\log_{2}\left(1+\frac{\tilde{p}_i\tilde{g}_i}{\tilde{I}_i+n}\right)\ge R_{\mathrm{min}},\\
	&\tilde{p}_i\le {p}_{\mathrm{max}}.
\end{align}
\end{subequations}

Similar to that in the IESH-s network, given the expected single-hop distance $r_0$, the optimal power allocation strategy can be obtained by MAPEL algorithm. Meanwhile, because the transport capacity of the network is concave with $r_0$, the optimal single-hop distance can be obtained by Golden Section Search Method. The detailed algortihm is presented in Algorithm \ref{IESH_g}.  

Based on Algorithm \ref{IESH_g}, we have obtained the transport capacity for the IESH-g network. The average rate for multi-hop network with small-scale fading factor considered can then be obtained by Theorem \ref{theorem2}. 
\section{Numerical Results}
\label{NR}
In this section, numerical simulations are presented to validate the accuracy the theoretical results. The values of some parameters used throughout this section are listed in the following. In the simulation, the noise power $n$ is set to $n= 10$mW and the power attenuation factor $\alpha$ is set to $\alpha=3$. Because of the imperfect transmitter-side channel state information in practice, we set the channel gain between the users by quantizing a corresponding Rayleigh random variable $\hbar$, the PDF of which is given by $f_{\bar{H}}(\hbar)=\hbar e^{\frac{-\hbar^2}{2}}$. In particular, we set $N_h=4$ and $\beta_b=0.25$ for $b\in\left\{1,2,3,4\right\}$ and the Rayleigh random variable is partitioned into four intervals given by $[0, 0.7585), [0.7585, 1.1774), [1.1774, 1.6651)$ and  $[1.6651, \infty)$. The Rayleigh random variables with values falling into each interval are quantized as the centroid of the interval. The small scale fading factor is obtained by $h=\hbar^2$. As a result, we set $[h_1 = 4.6045, h_2 = 1.9805, h_3 = 0.9392, h_4 = 0.2412]$. The partition of distance and maximum interference distance are set to $d_0=1$m and ${N^I_m}=10$, respectively. The unit of power is mW in the simulation. Each simulation runs $10^5$ times. 

\begin{figure}[t]
\centerline{\includegraphics[width=8cm]{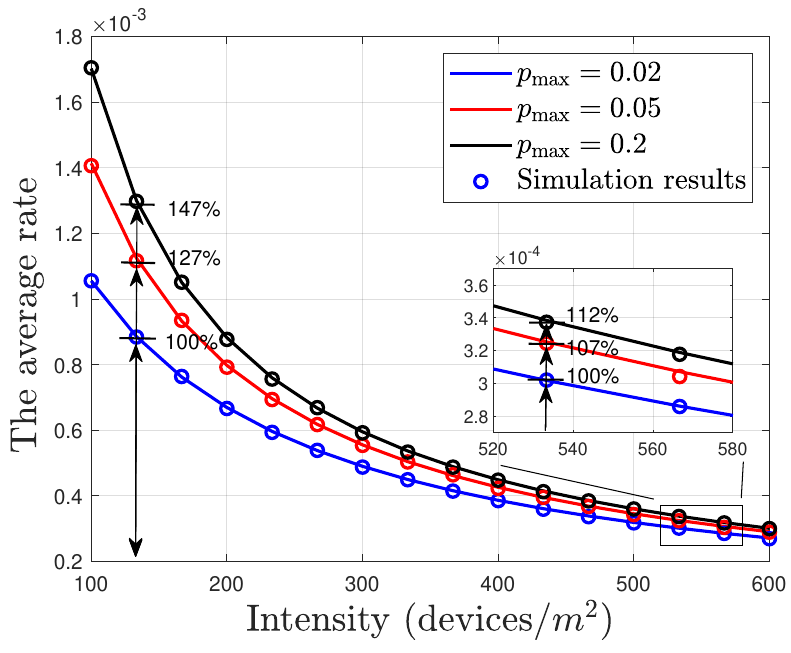}}
\caption{The average rate versus the intensity of the nodes of the massive communication network.}
\label{F1}
\end{figure}

Fig. \ref{F1} presents the average rate of the single-hop network under different nodes intensity $\lambda$. In Fig. \ref{F1}, we set ${N_m}=2$ , $N_b=2$ and $N_a=1$, respectively. We consider three different power constraints i.e. $p_{\mathrm{max}}=\left\{0.01, 0.02, 01\right\}$mW. From Fig. \ref{F1}, we can see that the simulation result and the results derived by the mean-field approximation match well with each other, which demonstrate the effectiveness of the mean-field approximation. As shown in Fig. \ref{F1}, the average rate of the network decreases with the intensity of devices. This is due to the fact that the total interference increases as the number of devices increases, leading to the reduced average rate. Fig \ref{F1} also shows that as the device intensity grows, the increase in average rate brought by the increase in power constraint becomes smaller. This is because the increase of transmission power will increase the mutual interference in the network in case that the device intensity is large.
\begin{figure}[t]
\centerline{\includegraphics[width=8cm]{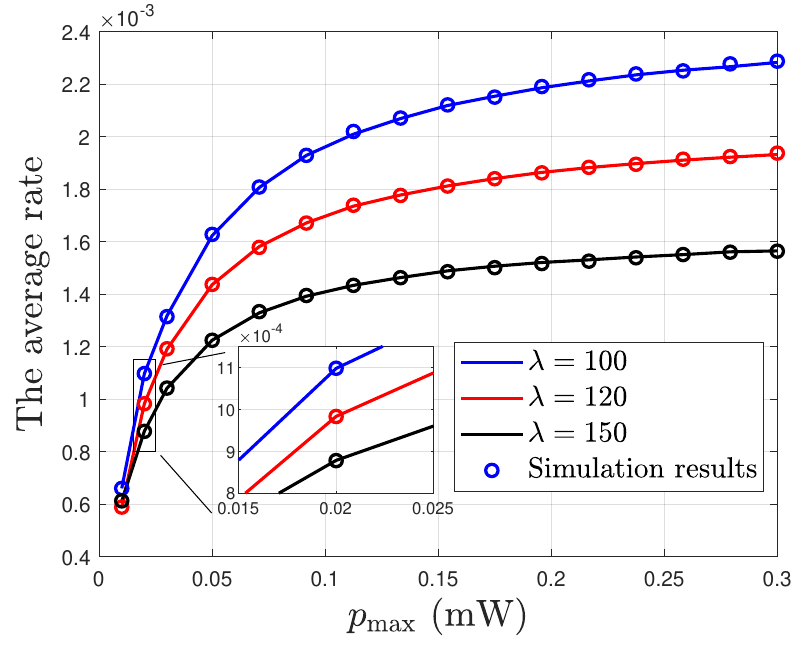}}
\caption{The average rate versus the maximum power constraint of the massive communication network.}
\label{F2}
\end{figure}

The average rate of the single-hop network under different power constraints is illustrated in Fig. \ref{F2}. It is observed that 
in case that the intensity of the devices is large, the increase of power constraint can not improve the average rate of the system. This is because when increasing the allocated transmission power of some group of devices, for each device in the group, the mutual interference from the same group will also increases.

\begin{figure}[t]
\centerline{\includegraphics[width=8cm]{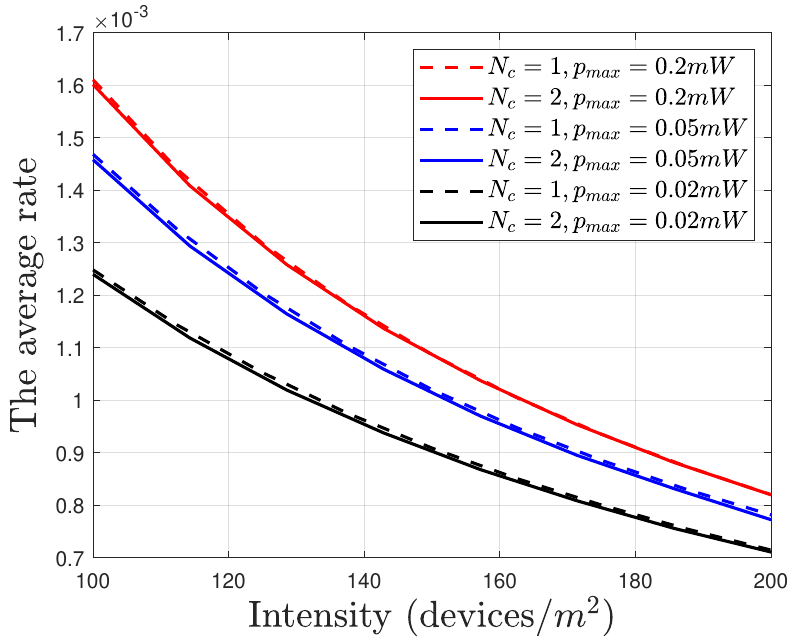}}
\caption{The comparison of the average rate under different $N_c$ for the massive communication network.}
\label{F8}
\end{figure}
The average rate of massive communication networks under different $N_c$ is illustrate in Fig. \ref{F8}. In Fig. \ref{F8} we compare two different $N_c$ values i.e., $N_c=1$ and $N_c=2$ under three different power constraints i.e., $p_{\mathrm{max}}=\{0.02, 0.05, 0.2\}$mW. From Fig. \ref{F8} we can see that increasing the number of partition intervals for the number of interfering nodes can not improve the average rate of the networks. This is because in massive communication networks, the intensity of the nodes is sufficiently large so that the interference power converges to the mean value according to the law of large numbers. This result indicates that in massive communication networks, only the CSI between the source and the destination nodes is helpful to improve the network spectrum efficiency. Based on this, a distributed power control strategy is feasible for practical large-scale networks. 

\begin{figure}[t]
\centerline{\includegraphics[width=8cm]{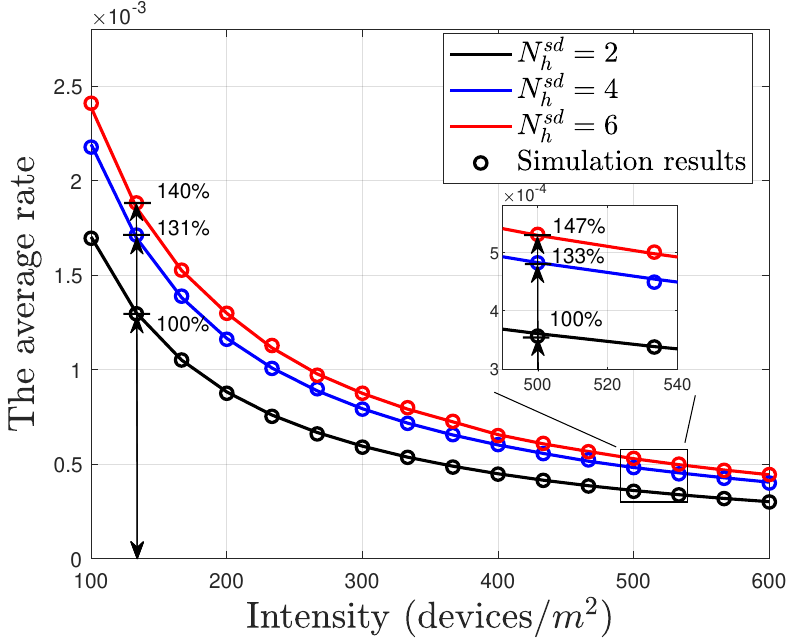}}
\caption{The comparison of the average rate under different $N_h^{sd}$ for the massive communication network.}
\label{F3}
\end{figure}
Fig. \ref{F3} presents the average rate of the massive communication network under different channel quantification of the channel state between the source and the destination nodes. In Fig. \ref{F3}, we consider three different channel quantification i.e. $N_h^{sd}=\{2,4,6\}$. From Fig. \ref{F3}, we can see that the accurate feedback of the channel between the source node and destination node can improve the spectrum efficiency of the network. This is because accurate CSI allows a more precise power allocation for users with different channel gain. However, the more accurate CSI means the more cost of channel estimation. The result in Fig. \ref{F3} also shows the tradeoff between the average rate of the network and the cost of channel estimation. 

\begin{figure}[t]
\centerline{\includegraphics[width=8cm]{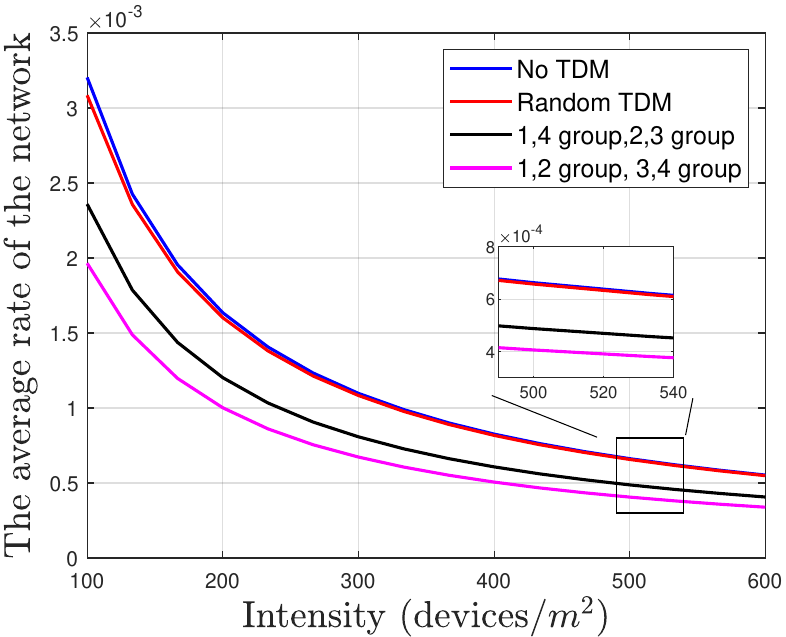}}
\caption{The comparison of different TDM schemes for the massive communication network.}
\label{F6}
\end{figure}
The average rate of the massive communication network under different TDM schemes is illustrated in Fig. \ref{F6}. In Fig. \ref{F6}, we consider four links with different channel gain where the channel state of each link is given by $h_1 = 2.1458, h_2 = 1.4073, h_3 = 0.9691, h_4 = 0.4911$. Moreover, four types of TDM schemes are considered, i.e. no time division, random time division, links 1 and 4 in one group and links 2 and 3 in one group, and links 1 and 2 in one group and links 3 and 4 in one group. From Fig. \ref{F6} we can see that the networks with no TDM and random TDM schemes have almost the same and the best average rate and the other two TDM networks have reduced average rate. This is because in the former two schemes, the power is allocated to the links with better channel state, which results in better spectrum efficiency and higher average rate. However, the latter two schemes assure fairness between the links. Fig. \ref{F6} shows the tradeoff between the average rate and the fairness of the networks.   

\begin{figure}[t]
\centerline{\includegraphics[width=8cm]{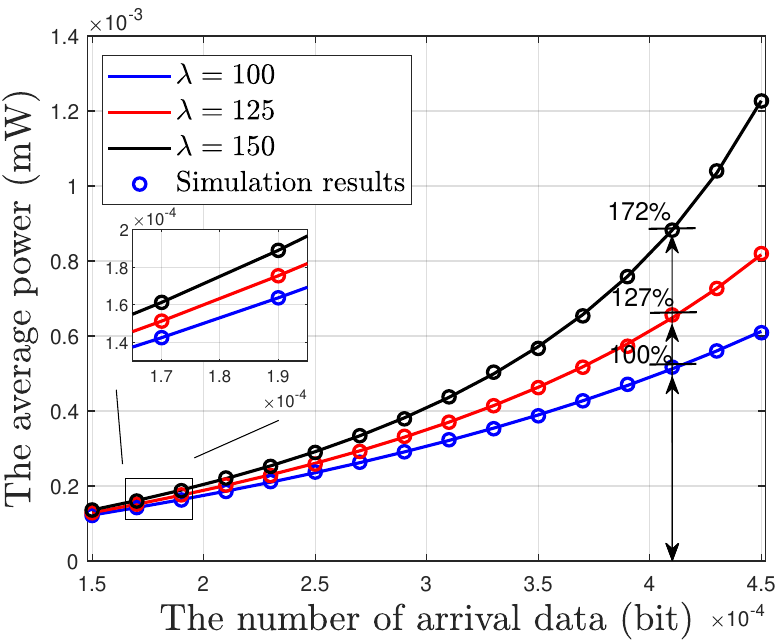}}
\caption{The average power versus the number of arrival data for the delay constrained massive communication network.}
\label{F9}
\end{figure}
The average power of the delay constrained massive communication network under different number of arrival data and node intensity is illustrated in Fig. \ref{F9}. In particular, we assume that the coherence time is much greater then the delay constraint. Therefore, the channel state for each link is fixed in each timeslot. The delay constraint and noise power are set to $T=3$ and $n=0.1$mW, respectively. At the start of each timeslot, the same number of data arrives at the buffer of each user. The channel gain between the users is given by $G=(1+d)^{-\alpha}$. The users are assumed to be PPP distributed. The maximum transmission distance is set to $3$m. From Fig. \ref{F9}, we can see that the simulation results and the results derived by the MFG, thereby demonstrating the effectiveness of MFG. Fig. \ref{F9} illustrates the basic tradeoff between the transmit power and the number of arrival bits. As shown in Fig. \ref{F9}, the average power increases fast with the number of arrival data. This is due to the fact that the increased transmission rate also increases the interference power between the users, leading to the decreased spectrum efficiency. Fig. \ref{F9} also shows that the increase in user intensity will induce the increase in the transmit power because of the higher interference.

\begin{figure}[t]
\centerline{\includegraphics[width=8cm]{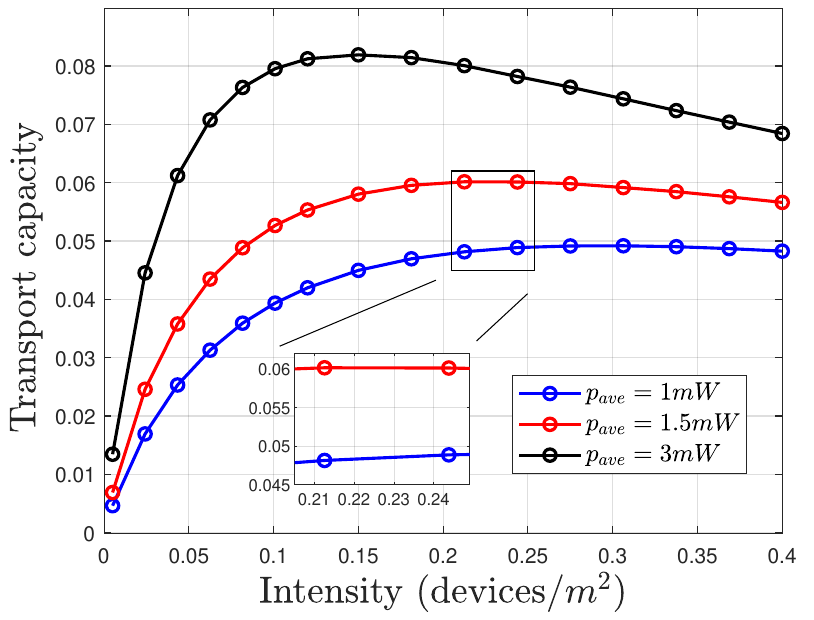}}
\caption{The transport capacity versus the device intensity for IESH-s netowork.}
\label{F4}
\end{figure} 
Fig. \ref{F4} shows the transport capacity of the IESH-s network under different user intensity and average power constraints. In Fig. \ref{F4}, we set $d_0=0.005$ and $N_{ds}=100$. From Fig. \ref{F4}, we can see that there exists an optimal device intensity to maximize the transport capacity of the IESH-s network, i.e. to maximize the average rate of infrastructure-free network without small-scale fading factor, which demonstrates the tradeoff between the routing efficiency and the interference. This is because the small device intensity results in high routing energy loss and the large device intensity increases the interference power. Fig. \ref{F4} also shows that the optimal devices intensity decreases with the increase of average power constraint, which means that more relay nodes are required to improve the throughput of infrastructure-free networks when the power supply is small.

\begin{figure}[t]
\centerline{\includegraphics[width=8cm]{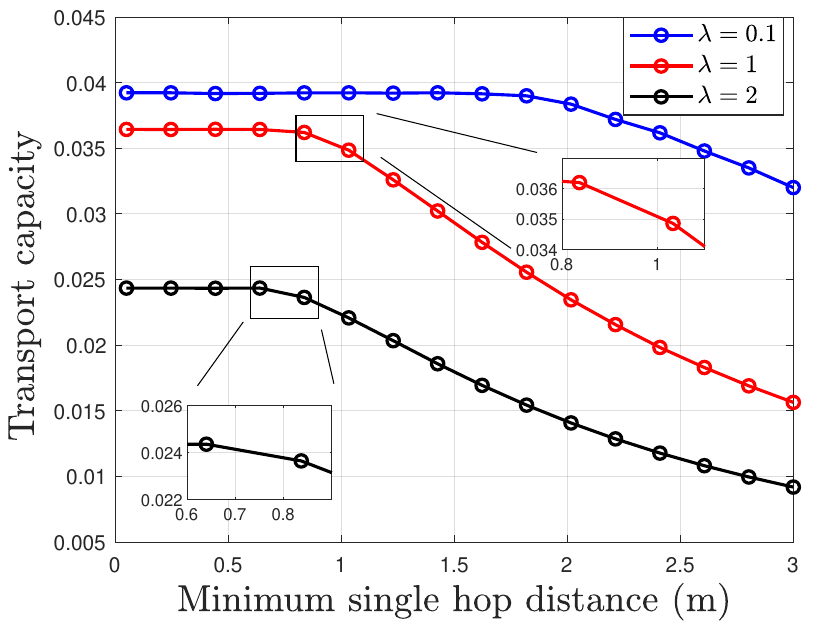}}
\caption{The transport capacity versus the minimum single-hop distance of the IESH-s network.}
\label{F5}
\end{figure}
In practical systems, because of the limited power supply, the total number of hops for each links is restricted. In Fig. \ref{F5}, we present the transport capacity of the IESH-s network under different minimum single-hop distance and different device intensity. From Fig. \ref{F5} we can see that the transport capacity decreases with minimum single-hop distance in case that $r_\mathrm{min}$ is large. This is due to the fact that the large transmission distance can increase the path loss, leading to the reduced transmission rate. Besides, Fig. \ref{F5} also shows that the optimal single-hop distance decreases with the user intensity. This is because in the OEPA strategy, the probability that the position of real relay nodes is close to the position of the desirable relay nodes increases with the user intensity. As a result, the optimal single-hop distance for the network decreases with the user intensity to reduce the path loss and improve the spectrum efficiency of the network.

\begin{figure}[t]
\centerline{\includegraphics[width=8cm]{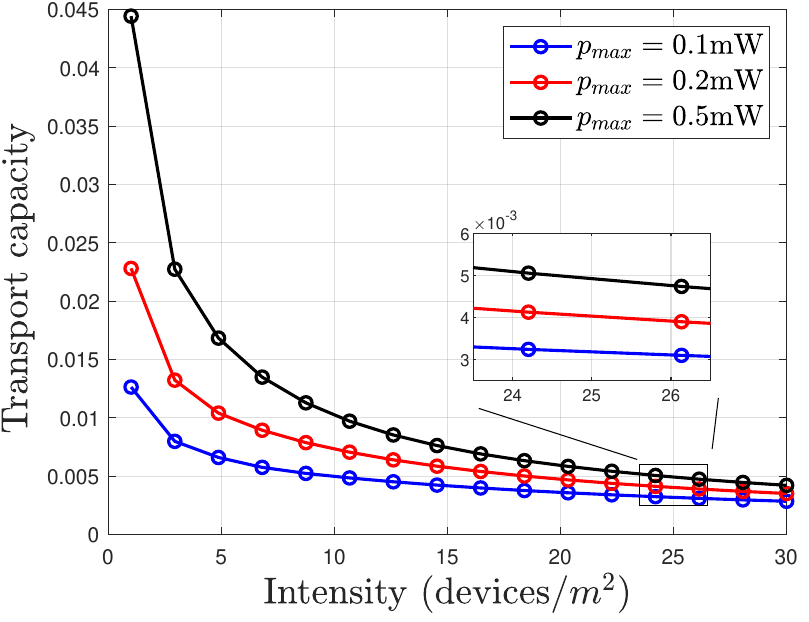}}
\caption{The transport capacity of the IESH-g network versus the intensity of the nodes in the network.}
\label{F7}
\end{figure}
In Fig. \ref{F7}, we present the transport capacity of the IESH-g network under different node intensity. From Fig. \ref{F7} we can see that the transport capacity of decreases with the intensity of the nodes in the network, which is different with the result in the IESH-s network. This is because in IESH-s network, the rate of the network is limited by the link with worst channel. When the intensity of the node is small, some links can not find the appropriate relay nodes. As a result, the rate of the network is decreased. However, in the IESH-g network, since the rate of the network is defined as the average value of all links, the power is allocated to the links with satisfactory channel states. Therefore, the links without appropriate relay nodes will not affect the throughout of the network. In contrast, the interference power grows with the node intensity of the network. As a result, the transport capacity of the IESH-g network decreases with the node intensity.    
\section{Conclusion}
\label{con}
To facilitate 6G standardization towards massive communications and ubiquitous connectivity, this paper  investigated the throughput of 6G networks with different spectrum sharing strategies and constraints. The mean-field approximation was exploited to decouple the complex mutual interference of massive networks and reduce the dimension of throughput maximizing problem. We proposed a tractable performance analysis approach that is independent of the size of the network for 6G networks. We further investigated the scenario with delay constraint by constructing a MFG problem. The effectiveness of channel orthogonalization for massive communication networks was discussed under different scenarios to facilitate system optimization. It was shown that the throughput of infrastructure-free network is proportional to the transport capacity of the corresponding single-hop network and the deployment of relay nodes can bring substantial energy and spectrum gain for infrastructure networks. Furthermore, only the CSI between the source and destination node of each link is required for spectrum sharing while the CSI of interfering links is unnecessary. The improvement brought by accurate CSI feedback from the users was also investigated. Numerical results demonstrated that the mean-field approximation method can efficiently estimate the performance of 6G networks without exponentially increasing computational complexity.

\bibliographystyle{IEEEtran}
\bibliography{ref}

\begin{thebibliography}{10}
\providecommand{\url}[1]{#1}
\csname url@samestyle\endcsname
\providecommand{\newblock}{\relax}
\providecommand{\bibinfo}[2]{#2}
\providecommand{\BIBentrySTDinterwordspacing}{\spaceskip=0pt\relax}
\providecommand{\BIBentryALTinterwordstretchfactor}{4}
\providecommand{\BIBentryALTinterwordspacing}{\spaceskip=\fontdimen2\font plus
\BIBentryALTinterwordstretchfactor\fontdimen3\font minus
  \fontdimen4\font\relax}
\providecommand{\BIBforeignlanguage}[2]{{%
\expandafter\ifx\csname l@#1\endcsname\relax
\typeout{** WARNING: IEEEtran.bst: No hyphenation pattern has been}%
\typeout{** loaded for the language `#1'. Using the pattern for}%
\typeout{** the default language instead.}%
\else
\language=\csname l@#1\endcsname
\fi
#2}}
\providecommand{\BIBdecl}{\relax}
\BIBdecl

\bibitem{6G_you}
C.-X. Wang \emph{et~al.}, ``On the road to 6{G}: Visions, requirements, key
  technologies, and testbeds,'' \emph{IEEE Communications Surveys \&
  Tutorials}, vol.~25, no.~2, pp. 905--974, Secondquarter 2023.

\bibitem{Hyper_connt}
H.~Lee \emph{et~al.}, ``Towards 6{G} hyper-connectivity: Vision, challenges,
  and key enabling technologies,'' \emph{Journal of Communications and
  Networks}, vol.~25, no.~3, pp. 344--354, Jun. 2023.

\bibitem{massive_capacity1}
X.~Chen and D.~Guo, ``Gaussian many-access channels: Definition and symmetric
  capacity,'' in \emph{IEEE Information Theory Workshop (ITW)}, Seville, Spain,
  Sep. 2013.

\bibitem{massive_capacity2}
X.~Chen and D.~Guo, ``Many-access channels: The gaussian case with random user
  activities,'' in \emph{IEEE International Symposium on Information Theory},
  Honolulu, HI, Jun. 2014.

\bibitem{massive_capacity3}
X.~Chen, T.-Y. Chen, and D.~Guo, ``Capacity of gaussian many-access channels,''
  \emph{IEEE Transactions on Information Theory}, vol.~63, no.~6, Jun. 2017.

\bibitem{massive_capacity_MIMO}
F.~Wei \emph{et~al.}, ``On the fundamental limits of {MIMO} massive multiple
  access channels,'' in \emph{IEEE International Conference on Communications
  (ICC)}, Shanghai, China, May 2019.

\bibitem{NOMA_massive}
Y.~Yuan \emph{et~al.}, ``{NOMA} for next-generation massive iot: Performance
  potential and technology directions,'' \emph{IEEE Communications Magazine},
  vol.~59, no.~7, pp. 115--121, Jul. 2021.

\bibitem{NOMA_poor}
M.~Vaezi \emph{et~al.}, ``Non-orthogonal multiple access: Common myths and
  critical questions,'' \emph{IEEE Wireless Communications}, vol.~26, no.~5,
  pp. 174--180, Oct. 2019.

\bibitem{NOMA_D}
Z.~Ding \emph{et~al.}, ``A survey on non-orthogonal multiple access for 5{G}
  networks: Research challenges and future trends,'' \emph{IEEE Journal on
  Selected Areas in Communications}, vol.~35, no.~10, pp. 2181--2195, Oct.
  2017.

\bibitem{PD_NOMA}
S.~M.~R. Islam \emph{et~al.}, ``Power-domain non-orthogonal multiple access
  ({NOMA}) in 5{G} systems: Potentials and challenges,'' \emph{IEEE
  Communications Surveys \& Tutorials}, vol.~19, no.~2, pp. 721--742,
  Secondquarter 2017.

\bibitem{cluster1}
X.~Chen and R.~Jia, ``Exploiting rateless coding for massive access,''
  \emph{IEEE Transactions on Vehicular Technology}, vol.~67, no.~11, pp.
  11\,253--11\,257, Nov. 2018.

\bibitem{cluster2}
Y.~Liu \emph{et~al.}, ``Fairness of user clustering in {MIMO} non-orthogonal
  multiple access systems,'' \emph{IEEE Communications Letters}, vol.~20,
  no.~7, pp. 1465--1468, Jul. 2016.

\bibitem{MMTC_D2D}
T.~Taleb and A.~Kunz, ``Machine type communications in 3gpp networks:
  potential, challenges, and solutions,'' \emph{IEEE Communications Magazine},
  vol.~50, no.~3, pp. 178--184, Mar. 2012.

\bibitem{D2D_scheduling}
M.~Sheng \emph{et~al.}, ``On-demand scheduling: achieving {QoS} differentiation
  for {D2D} communications,'' \emph{IEEE Communications Magazine}, vol.~53,
  no.~7, pp. 162--170, Jul. 2015.

\bibitem{subchannel}
M.-H. Han, B.-G. Kim, and J.-W. Lee, ``Subchannel and transmission mode
  scheduling for {D2D} communication in {OFDMA} networks,'' in \emph{2012 IEEE
  Vehicular Technology Conference (VTC Fall)}, Quebec City, QC, Canada, Sep.
  2012.

\bibitem{D2D_EE}
S.~Mumtaz \emph{et~al.}, ``Energy efficient interference-aware resource
  allocation in {LTE-D2D} communication,'' in \emph{IEEE International
  Conference on Communications (ICC)}, Sydney, NSW, Australia, Jun. 2014.

\bibitem{FlashLinQ}
X.~Wu \emph{et~al.}, ``Flashlin{Q}: A synchronous distributed scheduler for
  peer-to-peer ad hoc networks,'' \emph{IEEE/ACM Transactions on Networking},
  vol.~21, no.~4, pp. 1215--1228, Aug. 2013.

\bibitem{NTN_survey}
H.~Guo \emph{et~al.}, ``A survey on space-air-ground-sea integrated network
  security in 6{G},'' \emph{IEEE Communications Surveys \& Tutorials}, vol.~24,
  no.~1, pp. 53--87, Firstquarter 2022.

\bibitem{UAV_survey}
B.~Li, Z.~Fei, and Y.~Zhang, ``{UAV} communications for 5{G} and beyond: Recent
  advances and future trends,'' \emph{IEEE Internet of Things Journal}, vol.~6,
  no.~2, pp. 2241--2263, Apr. 2019.

\bibitem{UAV_BS}
M.~Mozaffari \emph{et~al.}, ``Wireless communication using unmanned aerial
  vehicles ({UAV}s): Optimal transport theory for hover time optimization,''
  \emph{IEEE Transactions on Wireless Communications}, vol.~16, no.~12, pp.
  8052--8066, Dec. 2017.

\bibitem{UAV_prompt}
Y.~Zeng, R.~Zhang, and T.~J. Lim, ``Wireless communications with unmanned
  aerial vehicles: opportunities and challenges,'' \emph{IEEE Communications
  Magazine}, vol.~54, no.~5, pp. 36--42, May 2016.

\bibitem{UAV_number}
H.~Zhao \emph{et~al.}, ``Deployment algorithms for {UAV} airborne networks
  toward on-demand coverage,'' \emph{IEEE Journal on Selected Areas in
  Communications}, vol.~36, no.~9, pp. 2015--2031, Sep. 2018.

\bibitem{UAV_number2}
J.~Lyu \emph{et~al.}, ``Placement optimization of {UAV}-mounted mobile base
  stations,'' \emph{IEEE Communications Letters}, vol.~21, no.~3, pp. 604--607,
  Mar. 2017.

\bibitem{UAV_cover}
M.~Mozaffari \emph{et~al.}, ``Efficient deployment of multiple unmanned aerial
  vehicles for optimal wireless coverage,'' \emph{IEEE Communications Letters},
  vol.~20, no.~8, pp. 1647--1650, Aug. 2016.

\bibitem{UAV_capacity}
V.~Sharma, M.~Bennis, and R.~Kumar, ``{UAV}-assisted heterogeneous networks for
  capacity enhancement,'' \emph{IEEE Communications Letters}, vol.~20, no.~6,
  pp. 1207--1210, Jun. 2016.

\bibitem{UAV_cover2}
A.~V. Savkin and H.~Huang, ``Deployment of unmanned aerial vehicle base
  stations for optimal quality of coverage,'' \emph{IEEE Wireless
  Communications Letters}, vol.~8, no.~1, pp. 321--324, Feb. 2019.

\bibitem{MAPEL_high_SINR}
D.~Julian \emph{et~al.}, ``{QoS} and fairness constrained convex optimization
  of resource allocation for wireless cellular and ad hoc networks,'' in
  \emph{Proceedings.Twenty-First Annual Joint Conference of the IEEE Computer
  and Communications Societies}, vol.~2, NY, USA, Jun. 2002.

\bibitem{capacity_gupta}
P.~Gupta and P.~Kumar, ``The capacity of wireless networks,'' \emph{IEEE
  Transactions on Information Theory}, vol.~46, no.~2, pp. 388--404, Mar. 2000.

\bibitem{MF}
P.~M. Chaikin and T.~C. Lubensky, \emph{Principles of Condensed Matter
  Physics}.\hskip 1em plus 0.5em minus 0.4em\relax Cambridge University Press,
  1995.

\bibitem{MF1}
``A class of mean field interaction models for computer and communication
  systems,'' \emph{Performance Evaluation}, vol.~65, no.~11, pp. 823--838,
  2008, performance Evaluation Methodologies and Tools: Selected Papers from
  ValueTools 2007.

\bibitem{MF2}
N.~Gast, B.~Gaujal, and J.-Y. Le~Boudec, ``Mean field for markov decision
  processes: From discrete to continuous optimization,'' \emph{IEEE
  Transactions on Automatic Control}, vol.~57, no.~9, pp. 2266--2280, Sep.
  2012.

\bibitem{MFG}
H.~Gao \emph{et~al.}, ``Energy-efficient velocity control for massive numbers
  of {UAV}s: A mean field game approach,'' \emph{IEEE Transactions on Vehicular
  Technology}, vol.~71, no.~6, pp. 6266--6278, June 2022.

\bibitem{MFG2}
D.~Wang \emph{et~al.}, ``Mean field game-based waveform precoding design for
  mobile crowd integrated sensing, communication, and computation systems,''
  \emph{IEEE Transactions on Wireless Communications}, vol.~23, no.~8, pp.
  10\,430--10\,444, Aug 2024.

\bibitem{MF_JSAC}
C.~Li, W.~Chen, and K.~B. Letaief, ``Achieving low latency in massive access: A
  mean-field approach,'' \emph{IEEE Journal on Selected Areas in
  Communications}, vol.~40, no.~5, pp. 1473--1488, May 2022.

\bibitem{MF_TWC}
C.~Li, W.~Chen, and K.~B. Letaief, ``Meeting hard delay constraint in massive
  access: A mean-field approach,'' \emph{IEEE Transactions on Wireless
  Communications}, vol.~23, no.~4, pp. 2961--2977, Apr. 2024.

\bibitem{MC_network}
M.~Chiang \emph{et~al.}, \emph{Power Control in Wireless Cellular Networks},
  2008.

\bibitem{MAPEL}
L.~P. Qian, Y.~J. Zhang, and J.~Huang, ``{MAPEL}: Achieving global optimality
  for a non-convex wireless power control problem,'' \emph{IEEE Transactions on
  Wireless Communications}, vol.~8, no.~3, pp. 1553--1563, Mar. 2009.

\end{thebibliography}
\end{document}